\documentstyle[11pt,aaspp4,amssym,psfig]{article}

\def\plotthree#1#2#3{\centering \leavevmode
\epsfxsize=.30\columnwidth \epsfbox{#1} \hfil
\epsfxsize=.30\columnwidth \epsfbox{#2} \hfil
\epsfxsize=.30\columnwidth \epsfbox{#3}}

\received{}
\accepted{}
\journalid{}{}
\articleid{}{}
\slugcomment{}
\lefthead{Kippen et al.\ }
\righthead{COMPTEL Gamma-Ray Burst Locations}

\begin{document}

\title{THE LOCATIONS OF GAMMA-RAY BURSTS MEASURED BY COMPTEL}

\author{\scshape
	R.~Marc~Kippen,\altaffilmark{1,2,3}
	James~M.~Ryan,\altaffilmark{1}
	Alanna~Connors,\altaffilmark{1}
	Dieter~H.~Hartmann,\altaffilmark{4}
	Christoph~Winkler,\altaffilmark{5}
	Lucien~Kuiper,\altaffilmark{6}
	Martin~Varendorff,\altaffilmark{7}\\
	Mark~L.~McConnell,\altaffilmark{1}
        Kevin~Hurley,\altaffilmark{8}
	Wim~Hermsen,\altaffilmark{6}
	and\\
	Volker~Sch\"onfelder\altaffilmark{7}
	}

\altaffiltext{1}{Space Science Center, University of New Hampshire, Durham, 
		NH 03824}

\altaffiltext{2}{Now at Center for Space Plasma and Aeronomic 
		Research, University of Alabama in Huntsville, AL 35899}

\altaffiltext{3}{Mailing address: ES-84, NASA/Marshall Space Flight
                 Center, Huntsville, AL 35812; marc.kippen@msfc.nasa.gov}

\altaffiltext{4}{Department of Physics and Astronomy, Clemson University, 
		Clemson, SC 29634}

\altaffiltext{5}{Astrophysics Division, ESA/ESTEC, NL-2200 AG Noordwijk, 
		The Netherlands}

\altaffiltext{6}{SRON-Utrecht, Sorbonnelaan 2, NL-3584 Utrecht, 
		The Netherlands}

\altaffiltext{7}{Max-Planck-Institut f\"ur extraterrestrische Physik, 
		Postfach 1603, D-85740 Garching, Germany}

\altaffiltext{8}{Space Sciences Laboratory, University of California, 
		Berkeley, CA 94720}


\centerline{\textit{Accepted for publication in the Astrophysical Journal,
1998 Jan. 1, Vol. 492}}

\begin{abstract}

The COMPTEL instrument on the {\it Compton Gamma Ray Observatory\/} is
used to measure the locations of gamma-ray bursts through direct
imaging of MeV photons.  In a comprehensive search, we have detected
and localized 29 bursts observed between 1991 April 19 and 1995 May
31.  The average location accuracy of these events is 1.25$\arcdeg$
(1$\sigma$), including a systematic error of $\sim$0.5$\arcdeg$, which
is verified through comparison with Interplanetary Network (IPN)
timing annuli.  The combination of COMPTEL and IPN measurements
results in locations for 26 of the bursts with an average ``error
box'' area of only $\sim$0.3 deg$^2$ (1$\sigma$).  We find that the
angular distribution of COMPTEL burst locations is consistent with
large-scale isotropy and that there is no statistically significant
evidence of small-angle auto-correlations.  We conclude that there is
no compelling evidence for burst repetition since no more than two of
the events (or $\sim$7\% of the 29 bursts) could possibly have come
from the same source.  We also find that there is no significant
correlation between the burst locations and either Abell clusters of
galaxies or radio-quiet quasars.  Agreement between individual COMPTEL
locations and IPN annuli places a lower limit of $\sim$100~AU (95\%
confidence) on the distance to the stronger bursts.

\end{abstract}


\pagebreak

\section{Introduction}

Cosmic gamma-ray bursts (GRBs) are among the brightest, most plentiful
astrophysical phenomena in the observable gamma-ray sky, yet they
present unique observational challenges.  The greatest challenge is in
measuring burst locations accurately.  Because of this difficulty,
much of our current understanding has come from large-scale
statistical measures, where the accuracy of individual burst locations
is not as important as the total number of bursts.  The Burst and
Transient Source Experiment (BATSE) has now measured the locations
(accurate to $\sim$5$\arcdeg$, on average) and intensities of more
than 1700 GRBs---yielding evidence that the sources are highly
isotropic on the sky, yet radially inhomogeneous (e.g., Meegan et
al. 1992; Briggs et al.\ 1996; Pendleton et al.\ 1996).  These results
rule-out the once-favored galactic neutron star paradigm (e.g., Hurley
1986; Higdon \& Lingenfelter 1990), but still allow three vastly
different distance regimes: heliocentric (distance $D \lesssim
10^4$~AU), extended Galactic halo ($D \gtrsim 200$~kpc) and
cosmological ($D \sim 1$~Gpc).  The GRB mystery is thus far from being
solved, as the range of possible distances spans over 12 orders of
magnitude and the source objects remain elusive.  In this paper we use
the COMPTEL instrument aboard the {\it Compton Gamma Ray Observatory}
(CGRO) to measure GRB locations with improved accuracy.  A smaller
sample of more-accurate burst locations cannot improve upon
large-scale measures, but improved location accuracy is a
fundamental requirement for important studies of small-scale angular
clustering, correlations with known objects and searches for
counterpart objects at other wavelengths.  In addition, the
large-scale distribution of unique sub-populations could provide
important insights.

Recent investigation of small-scale angular clustering of GRB sources
has been hampered by the relative inaccuracy of locations measured
with BATSE.  Several statistical studies have yielded conflicting
results indicating the presence (Quashnock \& Lamb 1993; Wang \&
Lingenfelter 1993) and absence (Meegan et al.\ 1995; Brainerd et al.\
1995; Tegmark et al.\ 1996) of small-angle correlations among BATSE
burst locations.  The presence of small-scale angular correlations
would be important as it could indicate repeating GRB sources, or
unresolved clusters---both of which are difficult to explain in most
scenarios involving catastrophic destruction or mergers (e.g.,
M\'{e}sz\'{a}ros \& Rees 1992; Narayan, Paczy\'{n}ski \& Piran 1992;
White 1993; Katz 1993; Bickert \& Greiner 1993).  Earlier studies
found no evidence to indicate burst repetition on time scales shorter
than $\sim$10 yr, but a majority of the locations used in these
studies were also relatively inaccurate (Schaefer \& Cline 1985;
Atteia et al.\ 1987).

The investigation of cross-correlations between GRBs and catalogs of
known objects has also been limited by imprecise locations.  Most
studies have failed to reveal any significant association with any
Galactic or extragalactic source catalog (Howard et al.\ 1994;
Nemiroff, Marani \& Cebral 1994, Harrison, Webber \& McNamara 1995).
Recently, however, BATSE 3B burst locations have been reported to be
weakly correlated ($\sim$2--3$\sigma$ significance) with galaxy
clusters (Kolatt \& Piran 1996; Marani et al.\ 1997; but see Hurley
et al.\ 1997) and radio-quiet quasars (Schartel, Andernach \& Greiner
1997).  Such correlations might be expected at some level if GRBs lie
at cosmological distances characteristic of the galaxy samples, where
the brighter, closer bursts are expected to show the strongest signal.
However, no significant correlations with either galaxy clusters or
quasars were found in an earlier analysis of accurate locations of
bright GRBs, where one might expect a stronger signal (Webber et al.\
1995).

The most crucial need for accurate burst locations is in the
identification of counterpart objects at other wavelengths---a
potential key to solving the GRB mystery.  Unfortunately, quiescent
counterpart emission at any wavelength appears to be very weak, or
non-existent (see Vrba 1996).  The last remaining hope in this field
is to detect transient counterpart emission soon after the onset in
gamma rays---a particularly difficult challenge as it requires quick
and accurate GRB locations.  Most simultaneous and near-simultaneous
measurements that have been obtained are relatively insensitive, but
indicate that if low-energy (i.e., non gamma-ray) counterpart emission
exists at all, it must be weak and/or short-lived ($< 1$ day).
Accurate and timely GRB localization is thus crucial for future
counterpart search efforts (see McNamara, Harrison \& Williams 1995).
The recent discovery of weak x-ray/optical emission from the
directions of GRB 970228 and GRB 970508 offers the most convincing
evidence obtained thus far that GRBs do produce transient emission at
other wavelengths (Costa et al.\ 1997a, 1997b; van Paradijs et al.\
1997; Bond 1997).  If these bursts are typical, then other
counterparts will be detected only through the use of sensitive
instruments combined with rapidly determined, accurate GRB locations.

The COMPTEL instrument has the unique capability of directly imaging
MeV photon sources over a wide ($\sim$1 sr, FWHM) field-of-view (FoV).
It is thus able to measure the locations and energy spectra of several
GRBs per year.  Full-sky exposure to bursts is achieved through long
observations, over several years, wherein the CGRO satellite is regularly
re-pointed.  In this paper we use the MeV imaging capability and
effective full-sky coverage of COMPTEL to accumulate a comprehensive
catalog of GRB locations observed during the first four years of the
CGRO mission.  The catalog constitutes a unique sample of relatively
accurate locations that is used to investigate the GRB spatial
distribution---providing insight into the distance/source object
mystery.  The energy spectra and temporal properties of these events
have been discussed elsewhere (Kippen et al.\ 1996a, 1997 and
references therein), as have preliminary locations of some of the
bursts (see Table~1).

\section{Instrumentation and Fundamental Analysis}

\subsection{Instrument Characteristics}

COMPTEL is a wide-field, double-scatter Compton telescope capable of
imaging gamma-ray sources in the energy range 0.75--30 MeV.  It
consists of two planar layers of scintillation detector modules
separated by $\sim$1.5 m.  The upper (D1) layer incorporates seven low
Z, low density liquid (NE213A) detectors (total area 4188~cm$^2$),
while the lower (D2) layer is composed of 14 high Z, high density
NaI(Tl) modules (total area 8620~cm$^2$).  A {\it telescope event\/}
is defined as a coincident set of interactions in a single pair of D1,
D2 modules in anti-coincidence with a signal from any of four plastic
charged particle veto domes.  For each telescope event, the energy
deposits and interaction locations within each detector are measured
using the signals from photomultiplier tubes affixed to the detector
modules.  The interaction sequence from D1 to D2 is identified by a
time-of-flight (ToF) coincidence measurement and the absolute time of
each event is tagged with an accuracy of 125~$\mu$s.  Rejection of a
large fraction of neutron-induced background events is realized
through the measurement and discrimination of pulse-shape (PSD) in D1.

The COMPTEL design favors the ideal interaction wherein an incident
photon undergoes a single Compton scatter in D1, followed by complete
absorption of the scattered photon in D2.  For such ideal telescope
events, the measurement of energy deposits in D1 and D2 yields the
scattered electron energy and scattered photon energy, respectively,
while the measurement of interaction locations within the detectors
reveals the direction of the scattered photon.  Combined, the energy
and location information constrains the incident photon direction to
an annular ring or {\it event circle} on the sky.  The angular radius
$\bar{\varphi}$ of an event circle is determined by the kinematics of
Compton scattering:

\begin{equation}
  \label{eqcompton}
  \cos \bar{\varphi} = 1 - m_e c^2 \left( {1 \over E_2} - {1 \over {E_1 + E_2}}
  \right)\ ,
\end{equation}

\noindent where $m_e c^2$ is the electron rest energy and $E_1$ and
$E_2$ are the energy deposit measurements in D1 and D2, respectively.
Ambiguity in the arrival direction of individual photons is removed
through the statistical superposition of many events with different
scatter angles, thereby forming a unique source signature.
Complicating matters are non-ideal source events, such as those
arising from multiple Compton scatters or pair production in D1 or
incomplete energy absorption in the detectors, for which equation
(\ref{eqcompton}) is invalid.  Non-ideal events result in a
complicated source signature that must be accounted for in the
imaging analysis.  Selection criteria applied to event parameters
(i.e., $E_1$, $E_2$, $\bar{\varphi}$, ToF and PSD) during analysis can
significantly alter the effective energy range, sensitivity and FoV of
the instrument.  A complete description of COMPTEL and its operation
and performance was presented by Sch\"onfelder et al.\ (1993).

\subsection{Burst Detection Capabilities}

COMPTEL employs two independent operating modes for observing
transient phenomena such as gamma-ray bursts.  In parallel with the main
{\it telescope mode\/}, a secondary {\it burst mode\/} uses two of the
D2 modules to accumulate time-resolved spectra in the energy range
0.1--10 MeV (see Hanlon et al.\ 1994 and references therein).  In this
paper we are concerned only with the source localization capabilities
of the telescope mode observations.

In detecting and localizing gamma-ray bursts, COMPTEL is in most cases
a photon-limited device, whose sensitivity is governed by low
detection efficiency rather than background.  Sensitivity in this case
depends on burst fluence, with the minimum detectable fluence given by

\begin{equation}
  \label{eqsmin}
  S_{\rm min} = {{\overline{E} N_{\rm S}} \over A_{\rm eff}}\ ,
\end{equation}

\noindent where $\overline{E}$ is the spectrum-dependent mean photon
energy (0.75--30 MeV), $A_{\rm eff}$ is the effective detection area
for ideal events and $N_{\rm S}$ is the number of ideal source events
that are required to construct a meaningful image (Winkler et al.\
1986).  Requiring 20 ideal events from a normally incident $E^{-\rm
2}$ power law burst source ($A_{\rm eff}$ = 31 cm$^2$; $\overline{E}$
= 2.8 MeV) yields $S_{\rm min} = 4.5 \times 10^{-6}$ erg~cm$^{-2}$.
Burst sensitivity decreases as the zenith angle $\vartheta_{\rm c}$
between the telescope pointing axis and the source direction
increases.  This relation effectively limits the FoV for bursts to
$\vartheta_{\rm c} \lesssim 60\arcdeg$.  It is illustrated in
Figure~1a for several different power law energy spectra.  The strong
dependence of burst sensitivity on fluence (rather than, e.g., peak
flux) results in a significant bias against the detection of short
duration bursts, since short bursts typically do not result in enough
events for meaningful imaging (even though the instantaneous flux may
be large).

\begin{figure}[ht]
\centerline{\epsscale{0.5}\plotone{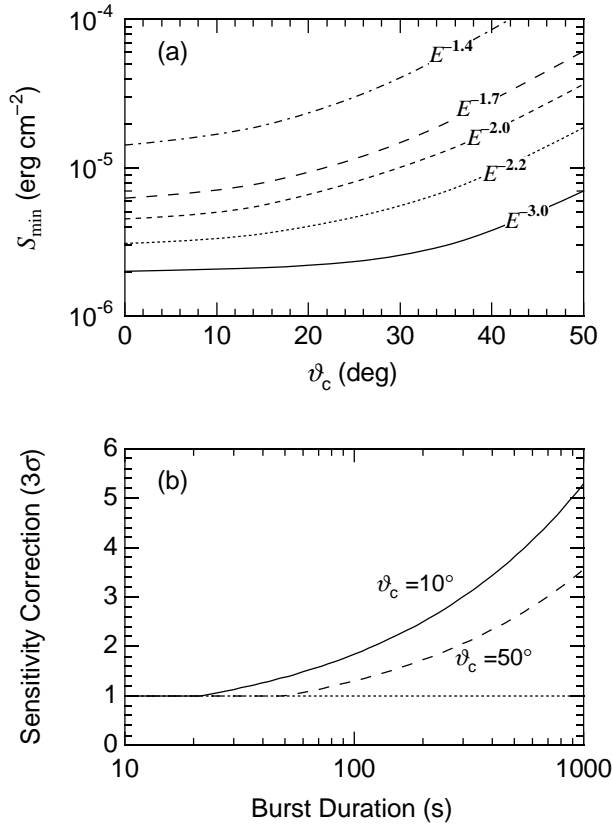}}
\medskip
\centerline{\parbox[]{14cm}{\figcaption[f1.eps]{
\baselineskip=9pt\footnotesize The estimated sensitivity of COMPTEL to
gamma-ray bursts.  (a) The minimum detectable fluence $S_{\rm min}$
(0.75--30 MeV) as a function of burst viewing angle $\vartheta_{\rm
c}$ for several different power law energy spectra, assuming zero
background.  For bursts longer than $\sim$25 s the estimates in (a)
must be multiplied by a background correction factor (b), which also
depends on the viewing angle, and can change by a factor of $\sim$3
due to orbital variations. \label{fig1} } } }

\end{figure}

The average rate of background events that could be confused with
ideal source events ranges from $\sim$0.2 Hz at $\vartheta_{\rm c} =
50\arcdeg$ to $\sim$0.6 Hz at $\vartheta_{\rm c} = 10\arcdeg$, with
orbital variations as large as a factor of three.  Background is thus
significant only in the detection of longer duration, low fluence
bursts.  The statistical detection significance for such bursts is

\begin{equation}
  \label{eqsdet}
  \sigma_{\rm det} \approx { N^{\prime}_{\rm S} \over 
   \sqrt{N^{\prime}_{\rm S} + 2 N_{\rm B}} }\ ,
\end{equation}

\noindent where $N^{\prime}_{\rm S}$ is the number of source counts
(related to the burst fluence through eq.\ [\ref{eqsmin}]) and $N_{\rm
B}$ is the number of background counts (roughly proportional to burst
duration).  When the number of source counts required for a
significant detection above background ($N^{\prime}_{\rm S}$) exceeds
the number of counts required for imaging ($N_{\rm S} \sim 20$),
detection is background-limited.  In this case, the sensitivity
estimates of Figure~1a are too low.  Correction factors (roughly
independent of spectral shape) for $\sigma_{\rm det} = 3$ are plotted
as a function of burst duration in Figure~1b.  The detection of bursts
shorter than $\sim$25 s is effectively always photon-limited
regardless of zenith angle.  For longer bursts, background becomes
increasingly important.  Fortunately, most long duration bursts
contain short, intense episodes whose detection is less dependent on
background.

For very intense bursts, or burst intervals, where fluence $S \gtrsim
10^{-4}$ erg~cm$^{-2}$, electronics and telemetry deadtime limit the
ability of COMPTEL to register and record events.  This affects the
detection of short, intense bursts, where significant numbers of
events fail to be registered and/or recorded.  Deadtime thus produces
an additional bias against the detection of short duration bursts.

\subsection{Burst Imaging Analysis}

The imaging analysis of COMPTEL data is performed in a
three-dimensional {\it dataspace\/} defined by the quantities ($\chi
,\psi,\bar{\varphi}$).  The coordinates ($\chi$,$\psi$) define the
direction of the vector between the D1, D2 interaction locations and
the scatter angle $\bar{\varphi}$ is computed from the energy deposit
measurements using equation (\ref{eqcompton}).  In this
representation, the point source signature, or {\it point spread
function\/} (PSF), is a partially filled cone with 45$\arcdeg$ opening
angle whose vertex is centered on the source position (Figure~2).  The
complicated diagonal shape of the cone mantle is determined by the
inherent measurement errors and the physics of non-ideal interactions,
that are all functions of source position and energy spectrum.
Dependence of the PSF shape on source position is weakened by
factoring-out a geometry term $\mathcal{G}$ (see Sch\"onfelder et al.\
1993).  This allows (to some degree of approximation) the use of a
single PSF for imaging sources located anywhere in the FoV.  The PSF
used for all burst source analysis in this paper has been derived from
Monte Carlo simulation (see Stacy et al.\ 1996) of the COMPTEL
response to a point source of gamma rays at $\vartheta_{\rm c} =
10\arcdeg$, with an assumed $E^{-2}$ power law energy spectrum.
Various standard selections are imposed on the event data in order to
reduce systematic effects due to incomplete knowledge of the PSF and
to improve the signal-to-noise ratio in the data (see Sch\"onfelder et
al.\ 1993).  These selections are applied equally to the PSF
simulations and the data.

\begin{figure}[ht]
\centerline{\epsscale{0.5}\plotone{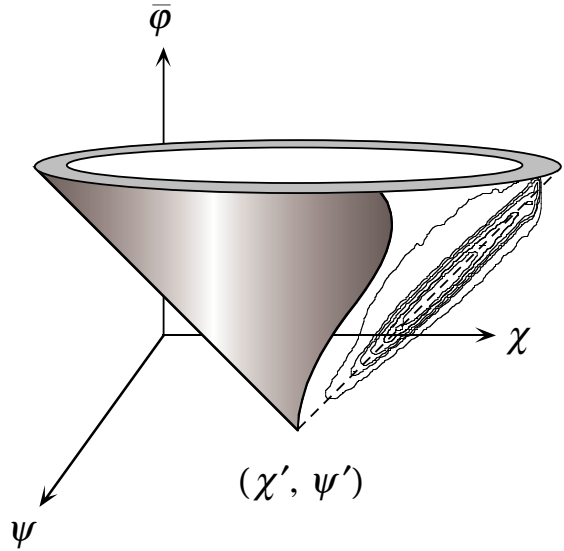}}
\centerline{\parbox[]{14cm}{
\figcaption[f2.eps]{\baselineskip=9pt\footnotesize Illustration of the
three-dimensional COMPTEL response to a point source located at sky
coordinates ($\chi'$,$\psi'$).  Contours show a cross-section of the
response function indicating the off-diagonal effects of intrinsic
measurement errors and non-ideal interactions. \label{fig2} } } }
\end{figure}

The imaging process amounts to searching for the PSF signature in the
binned ($\chi,\psi,\bar{\varphi}$) distribution of telescope events (a
mixture of source and background).  The bin-size is chosen to be on
the order of the instrumental resolution (1$\arcdeg$, 1$\arcdeg$ and
2$\arcdeg$ in $\chi$, $\psi$ and $\bar{\varphi}$, respectively),
resulting in few events per bin.  A maximum likelihood search
technique capable of handling the sparsely populated dataspace is
employed to obtain quantitative constraints on the source parameters.
The maximum likelihood method, described fully by de Boer et al.\
(1992), attempts to fit a point source (PSF) plus a model of the
background signature to the 3-d dataspace distribution of events
assuming Poisson counting statistics.  For a trial source position,
the likelihood of source plus background $\mathcal{L}$(S+B) and the
likelihood of the null hypothesis of pure background $\mathcal{L}$(B)
are independently maximized by allowing PSF and background scale
factors (but not shape) to vary.  A sky-map of the quantity

\begin{equation}
  \label{eqlrat} 
  \lambda = 2 \ln\left[\mathcal{L}{\rm (S+B)} \over
  \mathcal{L}{\rm (B)} \right]
\end{equation}

\noindent (the {\it maximum likelihood ratio\/} or MLR) is used to
estimate and constrain the source location and the statistical
significance of detection. The most probable source position
corresponds to the maximum $\lambda \equiv \hat{\lambda}$ in the
sky-map, which {\it formally\/} obeys the $\chi^2$ probability
distribution with three degrees of freedom (source intensity and
position).  Thus, $\hat{\lambda} \gtrsim 14.2$ indicates a
statistically significant 3$\sigma$ detection.  For such significant
detections the source location is constrained by the quantity
$\Delta\lambda = \hat{\lambda} - \lambda$, which is distributed as
$\chi^{2}_{2}$ so that formal 1$\sigma$, 2$\sigma$ and 3$\sigma$
statistical confidence regions in the sky-map are defined where
$\Delta\lambda \lesssim 2.3$, 6.2 and 11.8, respectively.

The formal relation between likelihood ratio and $\chi^2$ is based on
the implicit assumption of an accurate PSF and background model.  The
form of the background model is thus an important part of burst
imaging, even though the number of background events during bursts is
typically small.  The dataspace distribution of background events
present during GRBs is difficult to model due to complicated temporal
and spatial variations.  An empirical model provides the most
practical means of approximating this, but background count rates are
too low to provide a sufficient statistical sample of the full 3-d
distribution.  Fortunately, the ($\chi$,$\psi$) component of this
distribution is reasonably approximated by the geometry function
$\mathcal{G}$ mentioned above.  The $\bar{\varphi}$ component embodies
much of the background complexity. It is determined independently for
each burst by directly sampling events that were detected under
orbital history conditions similar to the burst observation.  In
practice, this is done by sampling events within a window of satellite
orbital parameters chosen to approximate the orbital history of CGRO
at the time of a burst, and thereby the background environment during
the burst. The $\bar{\varphi}$ distribution of directly sampled
background events is combined with the geometry function (properly
normalized) to form a 3-d background model that is then used in the
imaging process.  An example of real and modeled background is shown
in Figure~3.

\begin{figure}[ht]
\centerline{\epsscale{0.5}\plotone{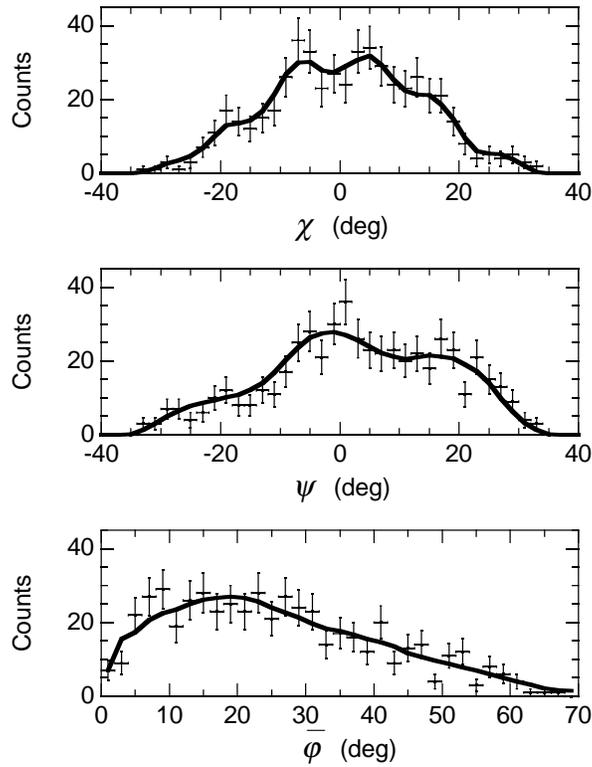}}
\centerline{\parbox[]{14cm}{
\figcaption[f3.eps]{\baselineskip=9pt\footnotesize An example of the
spatial distribution of COMPTEL background events during a gamma-ray
burst.  Observed ({\it data points\/}) and modeled ({\it solid
lines\/}) distributions are shown summed along each of the three
dataspace coordinates.  \label{fig3} } } }
\end{figure}

\subsection{Systematic Imaging Errors}

The approximations used in the imaging process are all potential
sources of systematic errors in the derived burst source parameters
(i.e., source significance and localization).  The major sources of
error are: a) the use of a single PSF for all bursts, regardless of
source location or energy spectrum; b) statistical fluctuations in the
PSF; c) the inherent accuracy of Monte Carlo simulations used to
compute the PSF; and d) the accuracy of the background model.  The
effects of each of these approximations have been studied with the use
of simulations like those used to determine the PSF.  Unfortunately,
since all combinations of potential causes of error cannot possibly be
simulated, this technique serves only as a predictive estimate of the
true errors.

Simulated sources with various, burst-like energy spectra, distributed
throughout the FoV and mixed with samples of real background were
subjected to the full imaging process with a variety of PSFs.  The
distribution of offsets between known source parameters and those
derived through imaging analysis of the simulated bursts yields an
estimate of systematic errors.  We find that approximation (a) has the
largest effect on source location accuracy, especially for
$\vartheta_{\rm c} \gtrsim 30\arcdeg$, but that the error is never
more than 0.75$\arcdeg$.  The other potential causes of location error
contribute only minimally.  Overall, simulated sources are correctly
localized by the MLR imaging technique to within an error of
0.5$\arcdeg$ ($\sim$90\% confidence).  We also find that
$\hat{\lambda}$ (indicating detection significance) is correctly
determined to within an error of approximately $\sim$30\%.  The
largest contributing cause in this case is approximation (d), since
the shape of the background in the observed COMPTEL dataspace is
difficult to model accurately.  The largest errors in $\hat{\lambda}$
are for sources located near the earth's limb, where atmospheric gamma
rays contribute most to the background and are not approximated well
by our model.  Bursts located near the earth's limb also result in
source photons backscattered off the atmosphere, which confuse the
true source signature.

\section{Observations}

\subsection{Burst Search Strategy}

To identify potential COMPTEL gamma-ray bursts, we rely on the BATSE
transient event triggering system.  This system automatically detects
statistically significant count rate (50--300 keV) increases above
background in the BATSE large area detectors (Fishman et al.\ 1989).
Triggers can occur from gamma-ray bursts, solar flares, magnetospheric
activity, Earth occultation of strong celestial sources and
terrestrial atmospheric events.  The COMPTEL burst search strategy is
to examine data only from the times of BATSE triggers that are
classified as GRBs by the BATSE analysis team (Mallozzi et al.\ 1993)
and whose BATSE locations satisfy $\vartheta_{\rm c} < 65\arcdeg$.
This strategy is effective in identifying COMPTEL GRB candidates.  An
exception occurs during BATSE data accumulation and readout periods
following all triggers (Meegan et al. 1996).  The BATSE trigger system
is disabled during these accumulation periods (usually lasting either
4~min or 8~min) and insensitive to new, weaker events during data
readout intervals (usually 90 min).  These gaps in coverage must be
taken into account when computing the exposure to bursts, as explained
in \S\ref{expsec}.

In the period between 1991 April 19 and 1995 May 31 there were 3608
BATSE triggers, 1297 of which have been classified as GRBs.  Of these,
397 have BATSE locations with $\vartheta_{\rm c} < 65\arcdeg$.
COMPTEL data are available for only 302 of the 397 burst candidates
due to gaps in telemetry caused by the failure of CGRO data recorders
in 1992.  Data around the time of each candidate trigger ($\pm$15 min)
have been manually examined to identify potential burst intervals.
Time intervals showing the most evidence for significant counts above
background were subjected to the maximum likelihood imaging process.
In cases where time intervals of significant emission were ambiguous,
the approximate full duration of the burst as measured by BATSE was
used for imaging.  MLR maps were computed over a grid of sky
coordinates with $1\arcdeg \times 1\arcdeg$ spacing and a final burst
detection threshold was applied to the images by requiring
$\hat{\lambda} > 20$---formally indicating a 3.8$\sigma$ point source
detection.  The possibility of systematic error in the determination
of $\hat{\lambda}$ is accounted for by using a conservative detection
threshold.  This ensures that only significant excesses
($\gtrsim$3$\sigma$) are accepted.  We expect, at the most, one
spurious detection in the search of 302 candidates.

\subsection{COMPTEL Gamma-Ray Burst Location Catalog}

Over the four years of observations, we have accumulated a catalog of
29 gamma-ray bursts that satisfy the $\hat{\lambda} > 20$ detection
requirement.  These bursts are summarized in Table~1. Each detection
is identified by the year ({\it yy\/}), month ({\it mm\/}) and day
({\it dd\/}) of the burst in the form GRB {\it yymmdd\/}.  The next
three columns give the number, Truncated Julian Day (TJD), and seconds
of the day of the corresponding BATSE burst trigger.  This is followed
by the value of $\hat{\lambda}$ and the most probable COMPTEL burst
location expressed in Spacecraft ($\vartheta_{\rm c}, \varphi_{\rm
c}$), Celestial ($\alpha, \delta$; epoch J2000.0) and Galactic ($l,
b$) coordinates.  Preliminary locations of many of the bursts have
been published earlier, as indicated.  This catalog includes two
bursts in addition to those reported in the preliminary analysis of
Kippen et al.\ (1996b).  One of these bursts (GRB~920627) was
mistakenly neglected in the preliminary search and the other
(GRB~950522) was beyond the time-interval of the earlier analysis.

\begin{table}[p]
\psrotatefirst
\psfig{figure=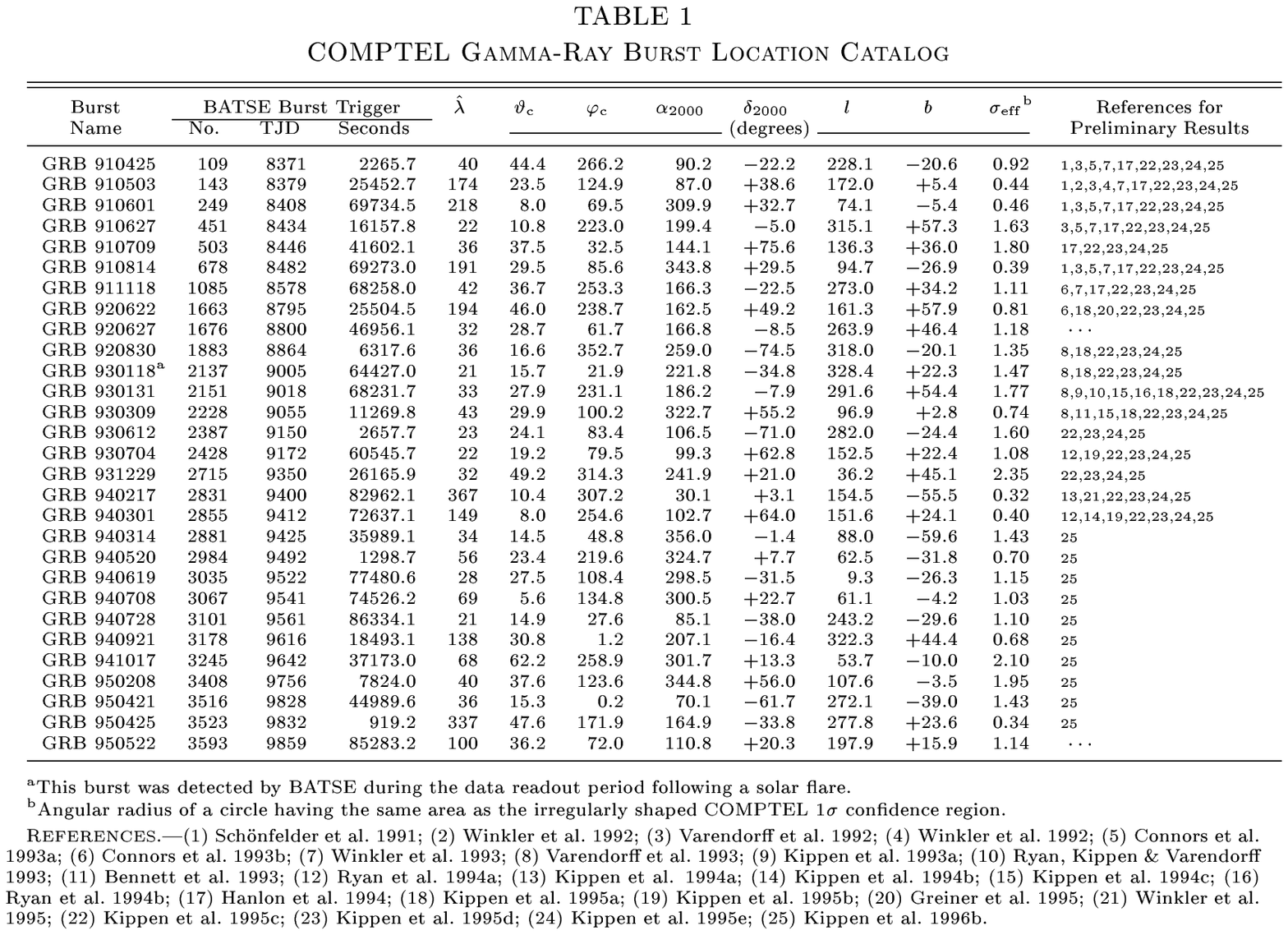,width=15cm,angle=90}
\label{tab1}
\end{table}

Statistical uncertainty regions around the most probable burst
locations have been estimated by computing smoothed (cubic spline)
contours of the MLR sky-maps at levels corresponding to 1$\sigma$,
2$\sigma$ and 3$\sigma$ confidence.  Celestial coordinate sky maps of
these contours are shown in Figure~4.  In general, the localizations
are irregularly shaped regions, making it difficult to compare
uncertainties between different bursts.  A convenient {\it
approximate\/} measure of the statistical location error is the radius
of a circle having the same area as the region enclosed by the MLR
contours.  The value of this quantity ($\sigma_{\rm eff}$) for the
1$\sigma$ confidence level is given in Table~1.  The statistical
location error is usually inversely related to the number of source
events.  Thus, the error radius of a high-fluence burst like
GRB~940217 ($\sigma_{\rm eff}=0.32\arcdeg$) is much smaller than that
of a weaker detection like GRB~910627 ($\sigma_{\rm
eff}=1.63\arcdeg$).  This behavior is complicated, however, by the
character of the off-axis telescope response, which introduces
additional uncertainty in the azimuthal ($\varphi_{\rm c}$) direction
for all bursts at large zenith angle ($\vartheta_{\rm c} \gtrsim
30\arcdeg$).  The average statistical location uncertainty, as
approximated by $\sigma_{\rm eff}$, for the full COMPTEL burst sample
is 1.13$\arcdeg$.  A circle with this angular radius is included in
each of the plots of Figure~4 for comparison.  Note that systematic
errors are not included in the MLR localizations.

\begin{figure}[p]
\epsscale{1.0} 
\plotthree{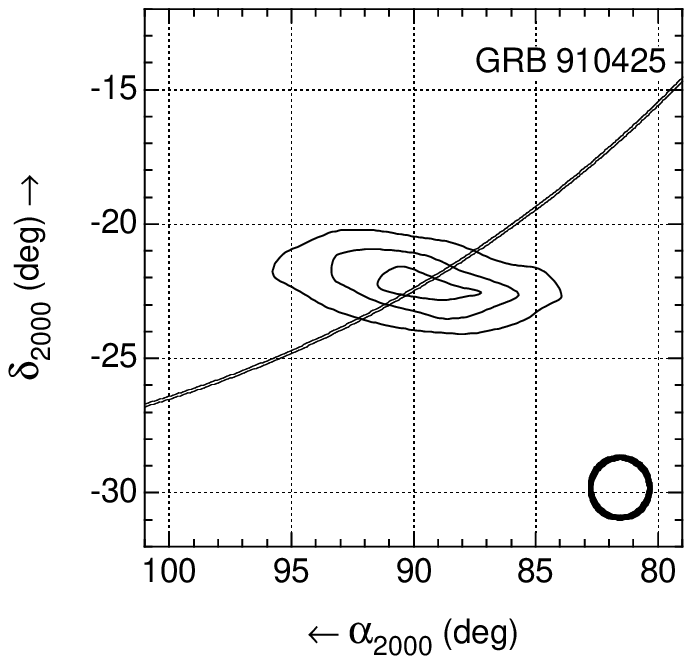}{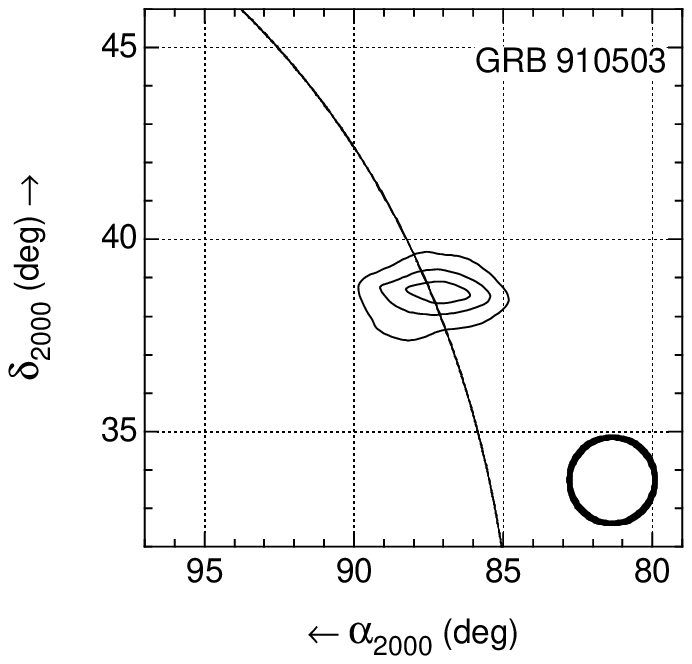}{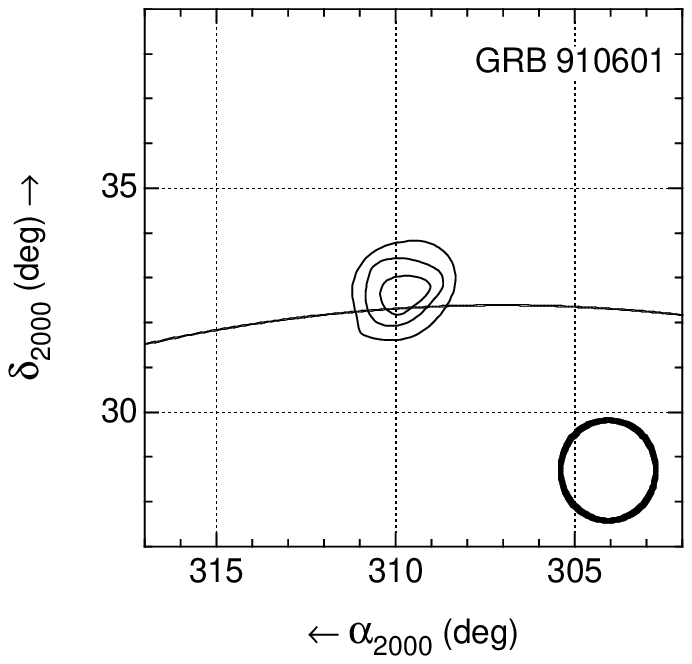} \\
\plotthree{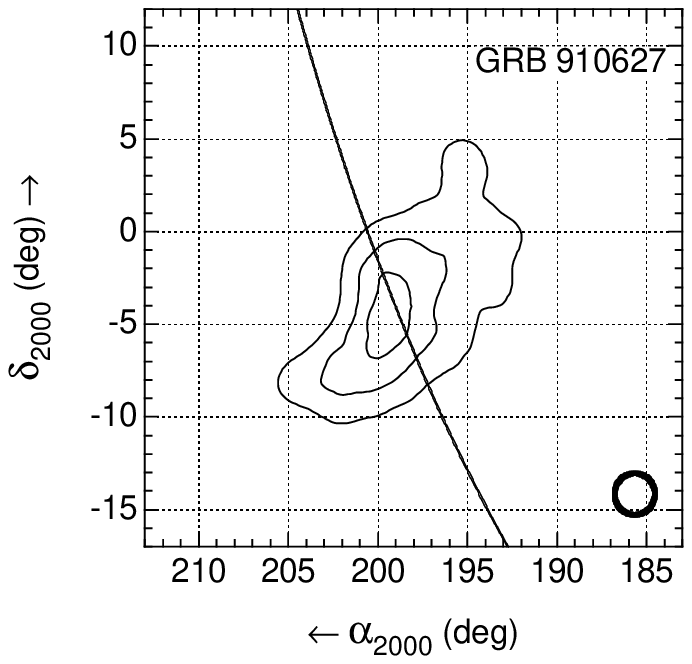}{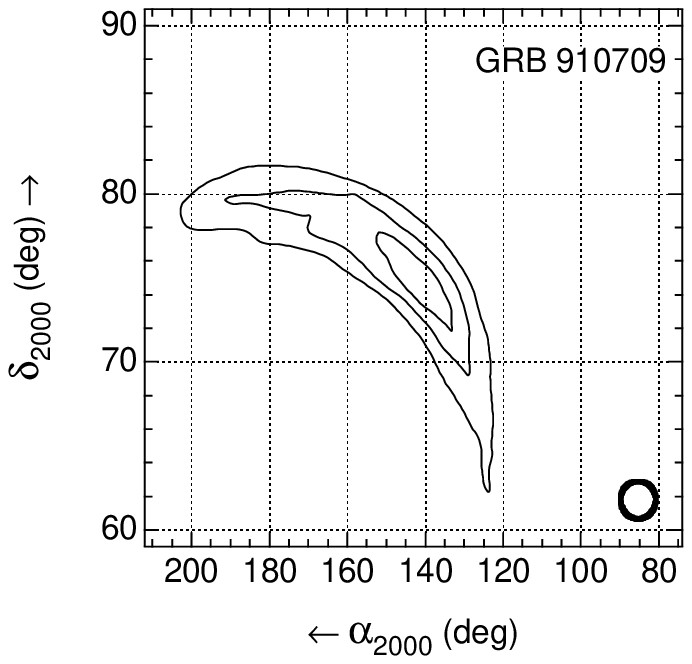}{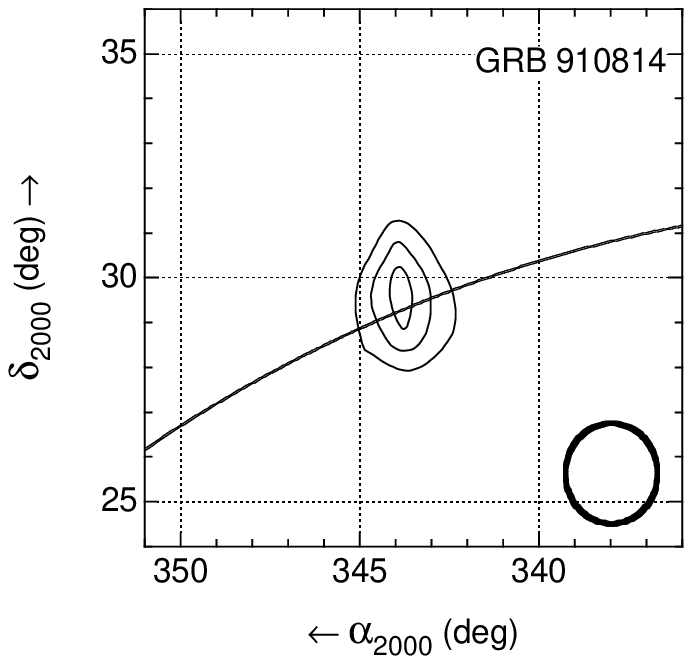} \\
\plotthree{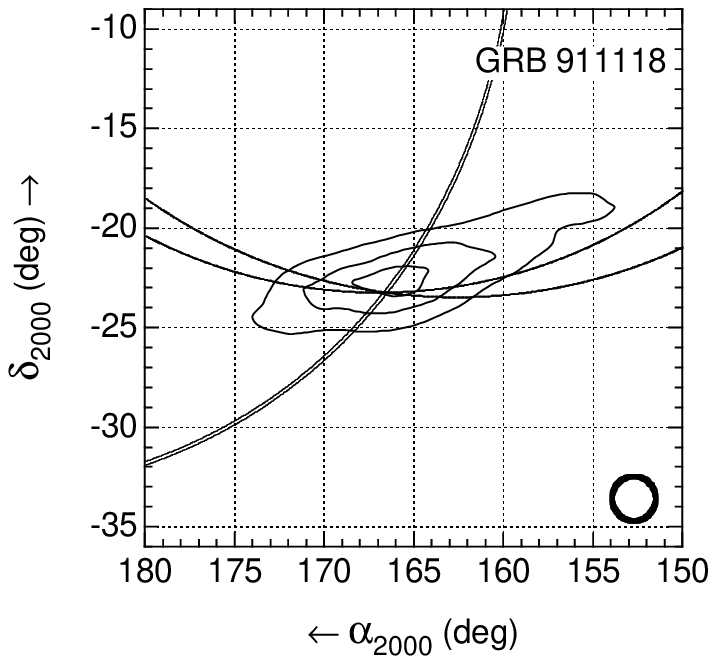}{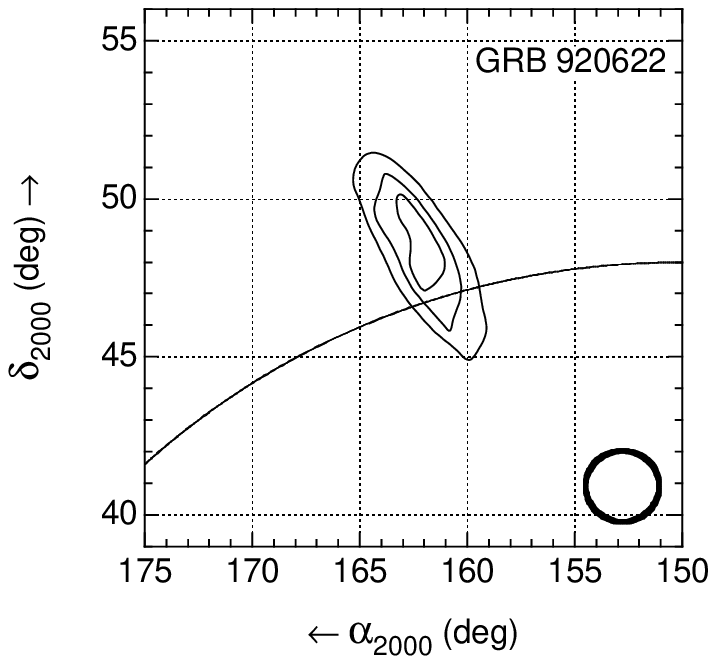}{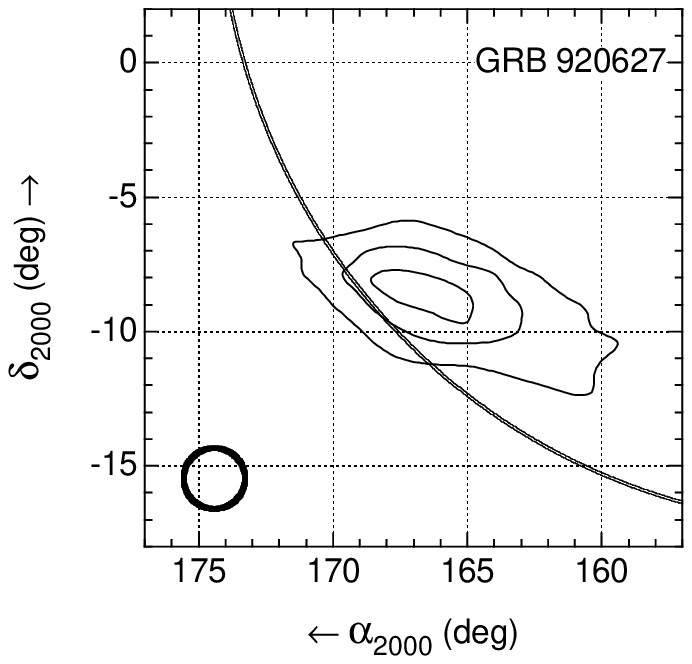}

\medskip
\centerline{\parbox[]{14cm}{
\figcaption[f4.eps]{\baselineskip=9pt\footnotesize These plots show
the locations of each of the COMPTEL bursts in Celestial coordinates
(epoch J2000.0).  Smoothed contours obtained through maximum
likelihood imaging enclose 1$\sigma$, 2$\sigma$ and 3$\sigma$
statistical confidence regions.  Where available, IPN timing annuli
(Hurley et al.\ 1996) are overlaid, with the 3$\sigma$ error in the
annulus width indicated by two concentric arcs.  For scale reference,
a circle corresponding to the average 1$\sigma$ area of all the bursts
(angular radius 1.13$\arcdeg$) is shown in each plot ({\it thick
line\/}). \label{fig4} } } }
\end{figure}

\setcounter{figure}{3}
\begin{figure}[p]
\epsscale{1.0} 
\plotthree{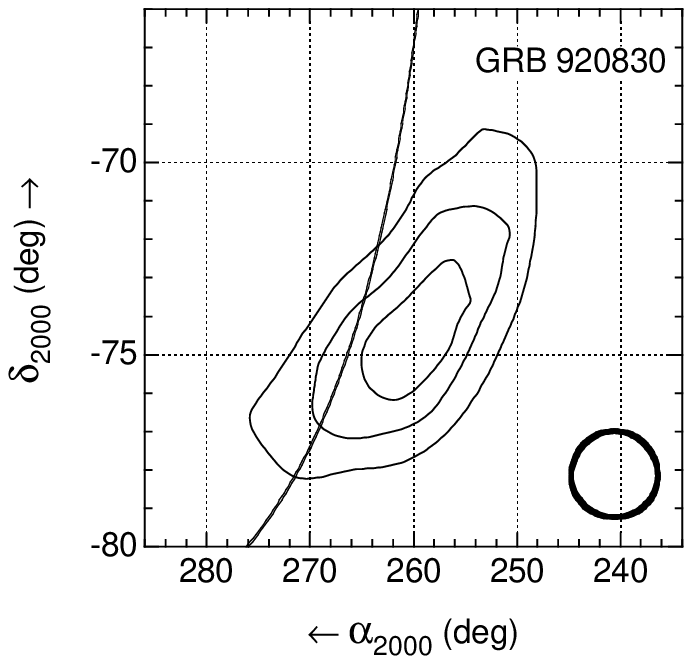}{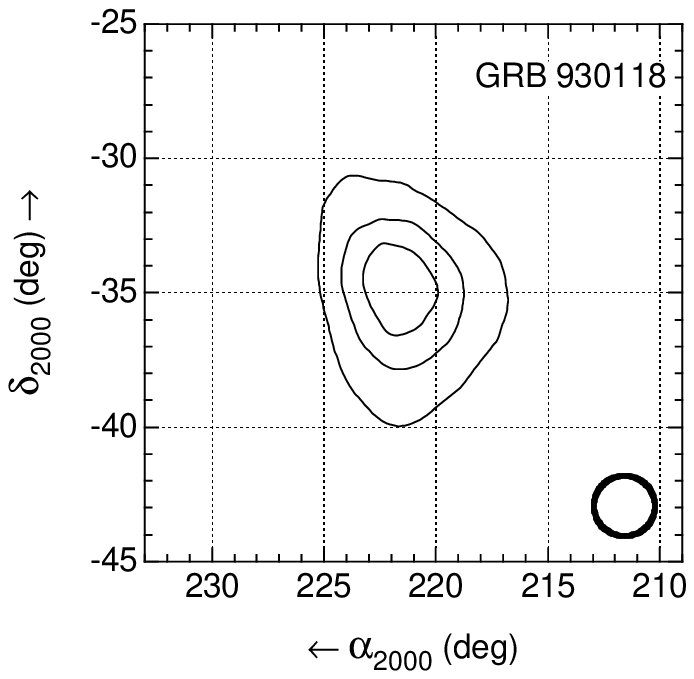}{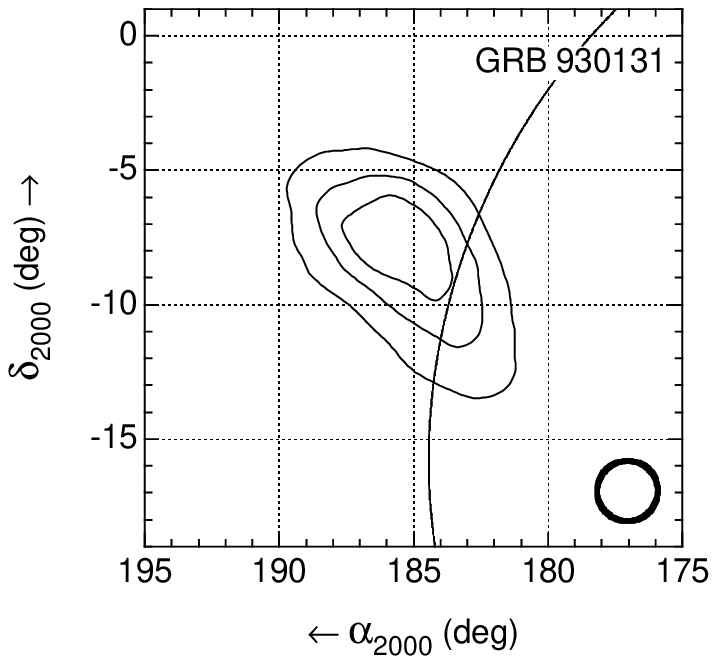} \\
\plotthree{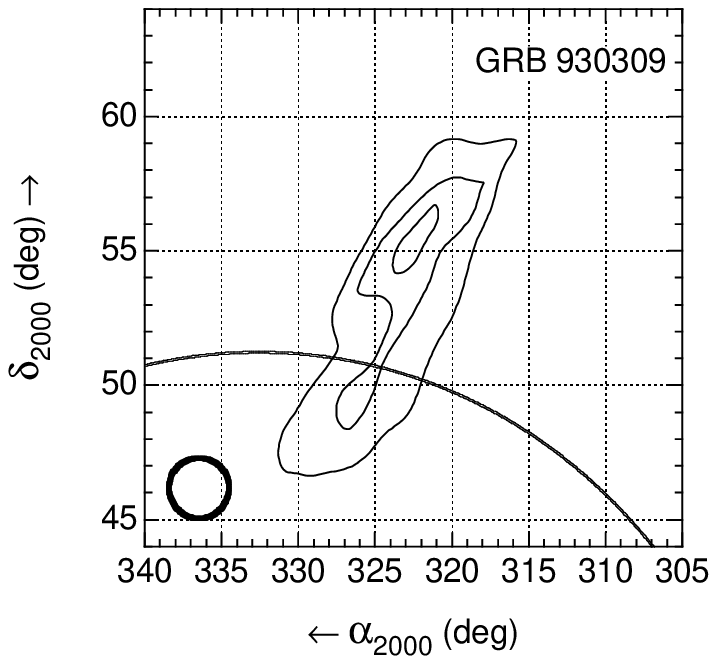}{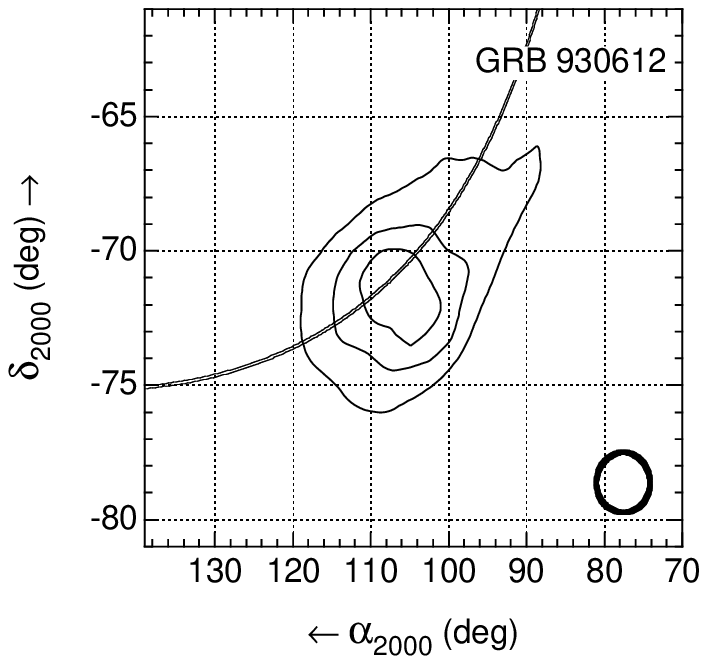}{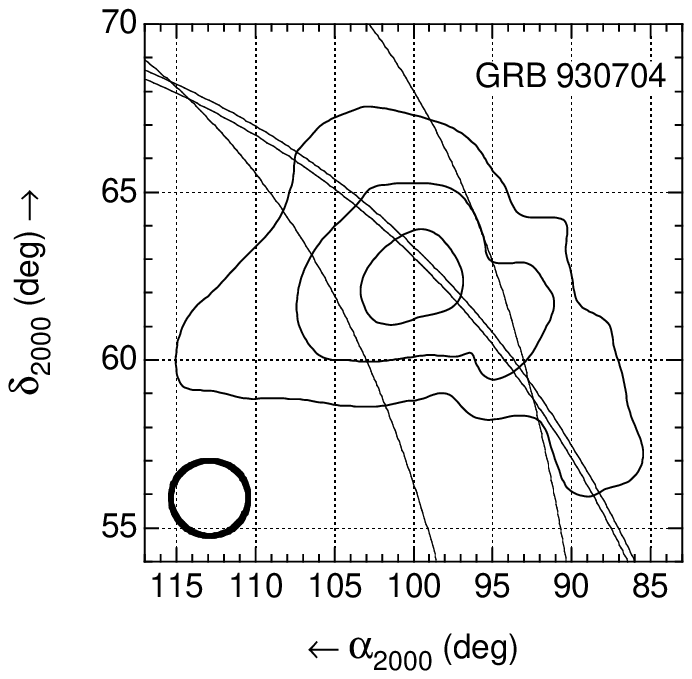} \\
\plotthree{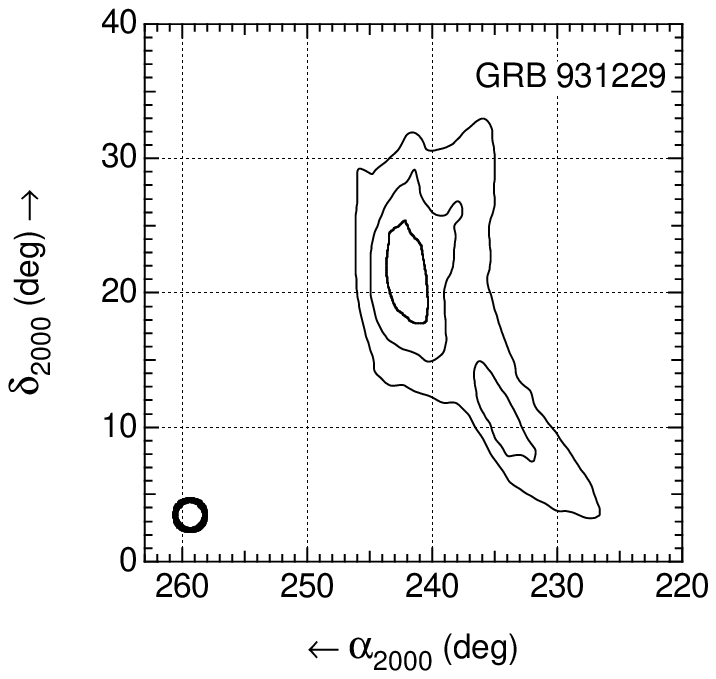}{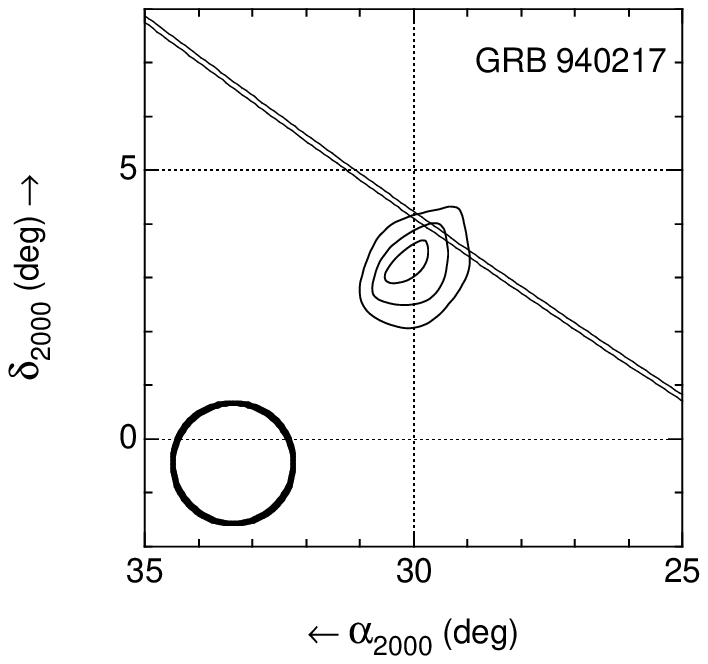}{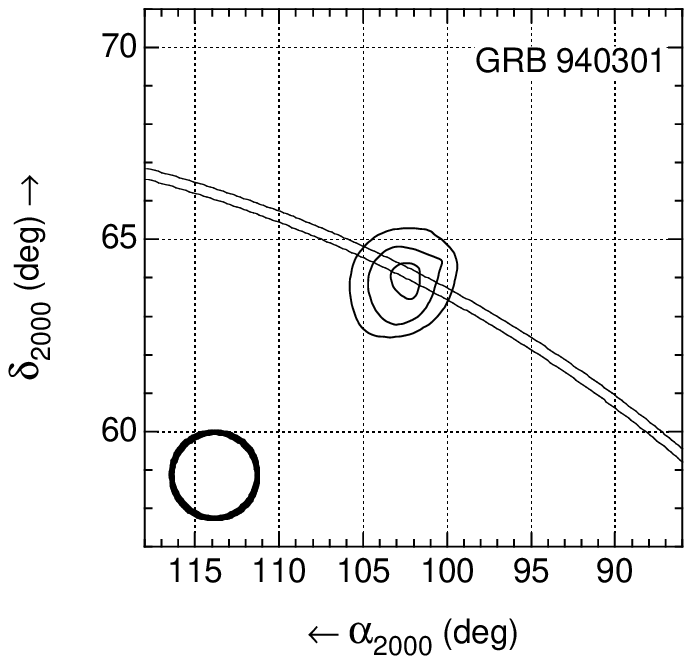} \\
\plotthree{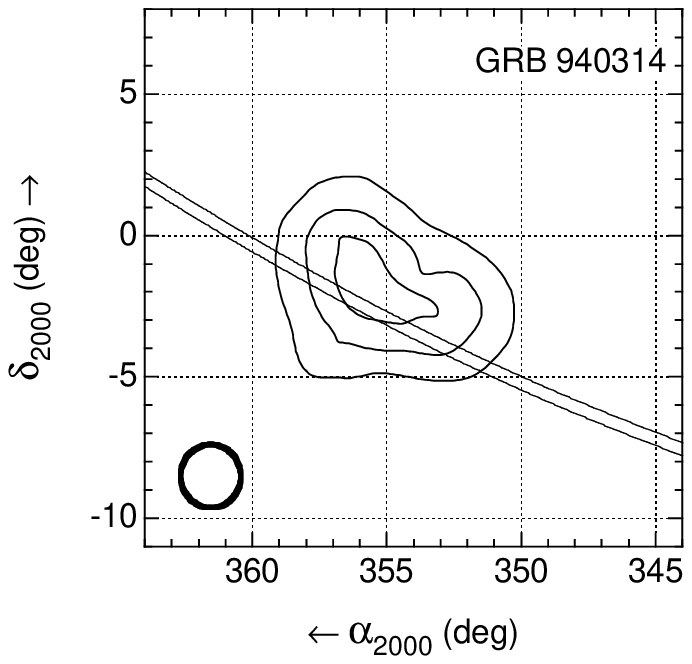}{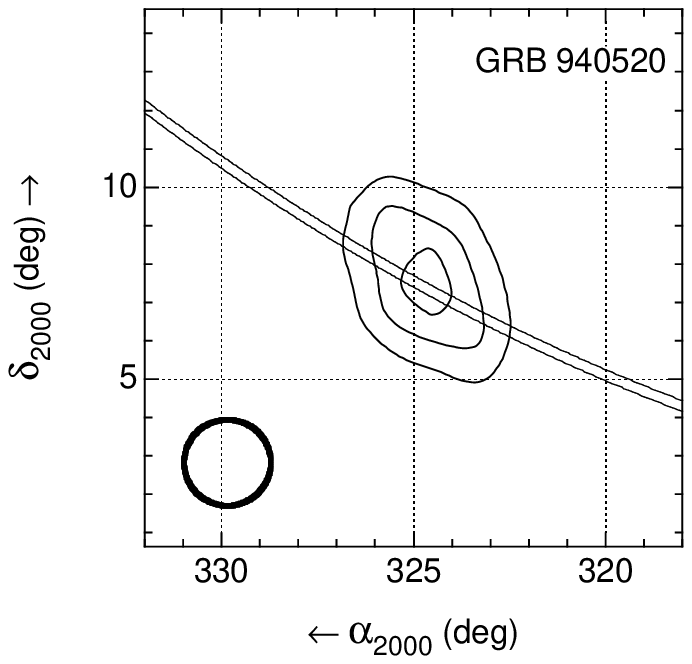}{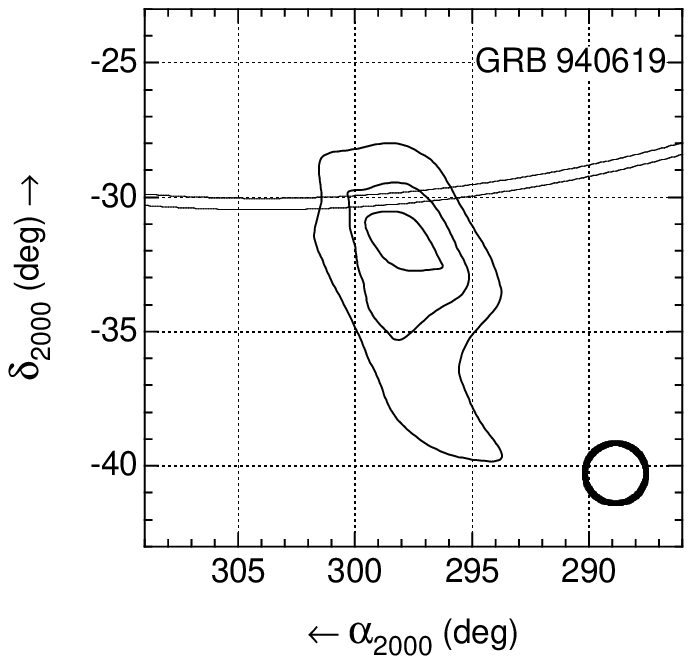}

\medskip
\centerline{\parbox[]{14cm}{
\figcaption[f4.eps]{\baselineskip=9pt\footnotesize Continued.} } }
\end{figure}

\setcounter{figure}{3}
\begin{figure}[tp]
\epsscale{1.0}
\plotthree{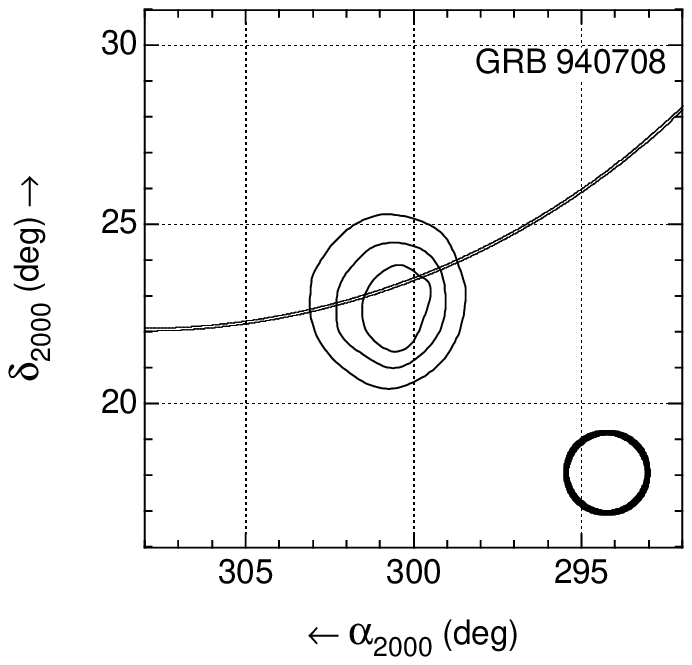}{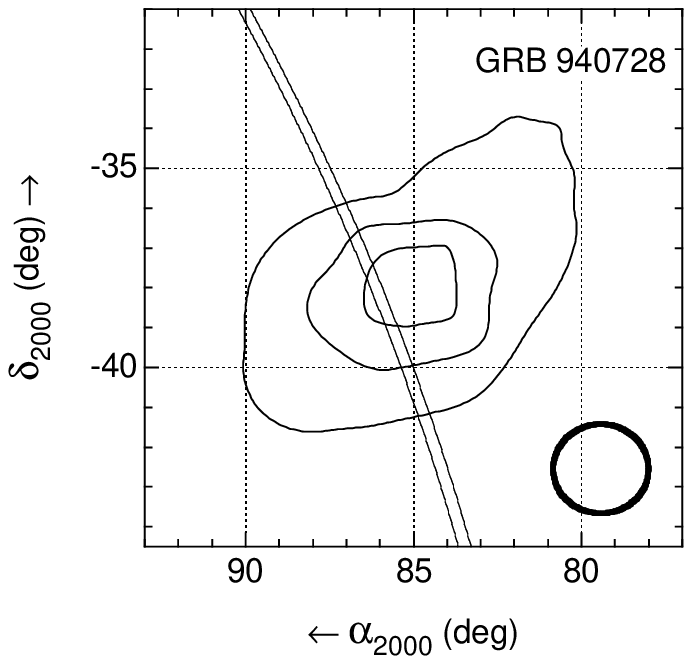}{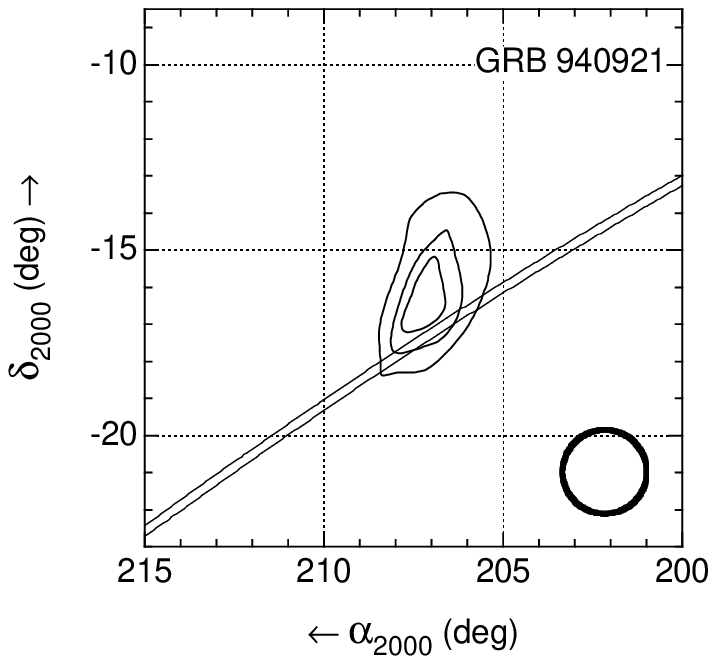} \\
\plotthree{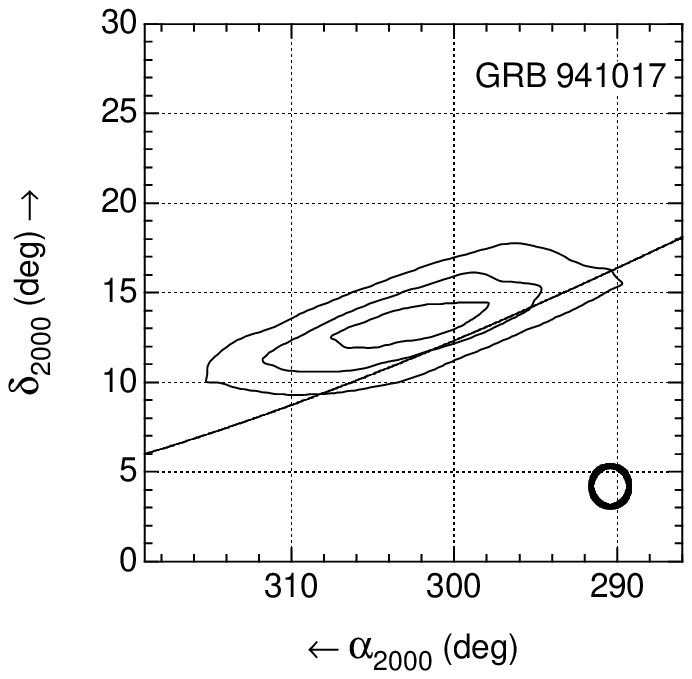}{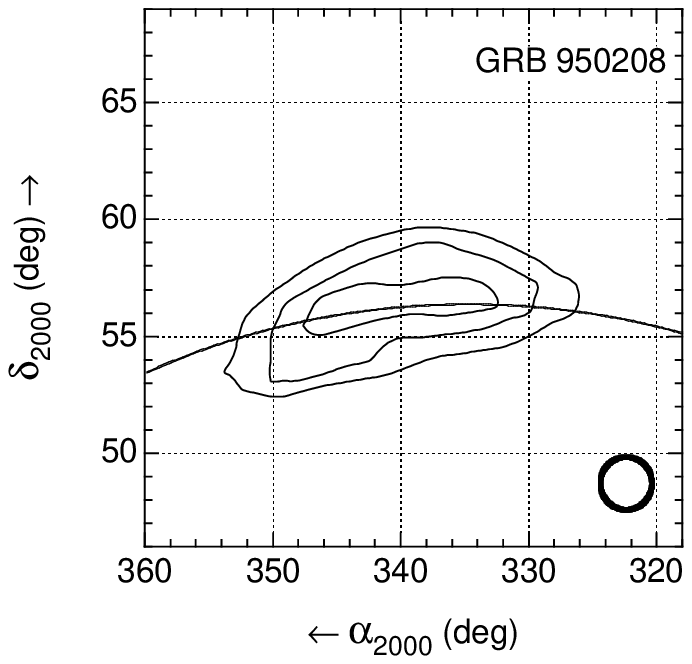}{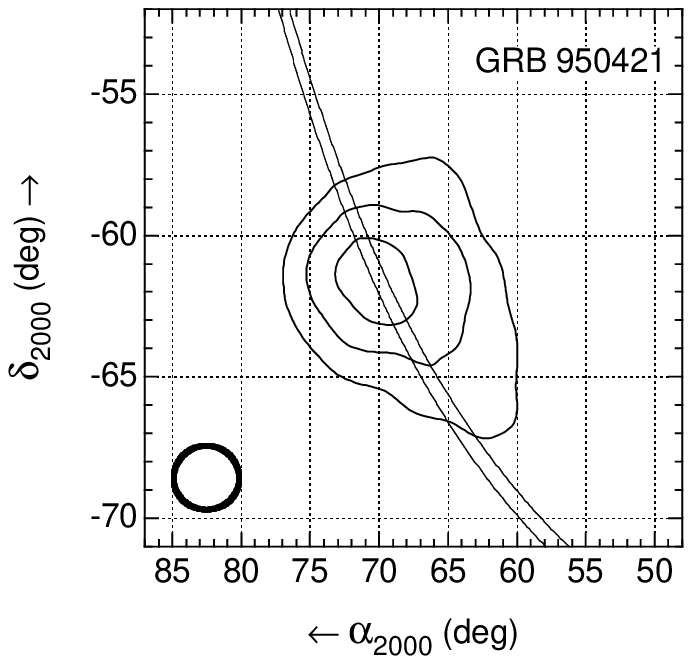} \\
\plotthree{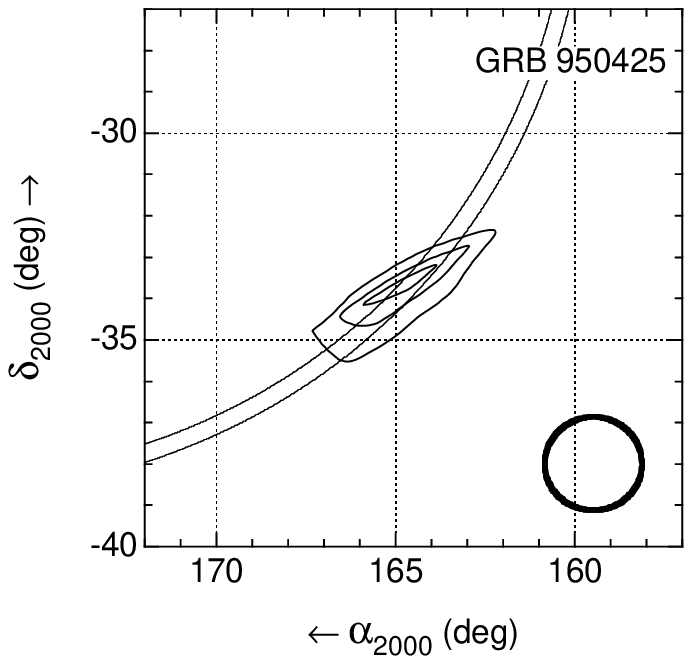}{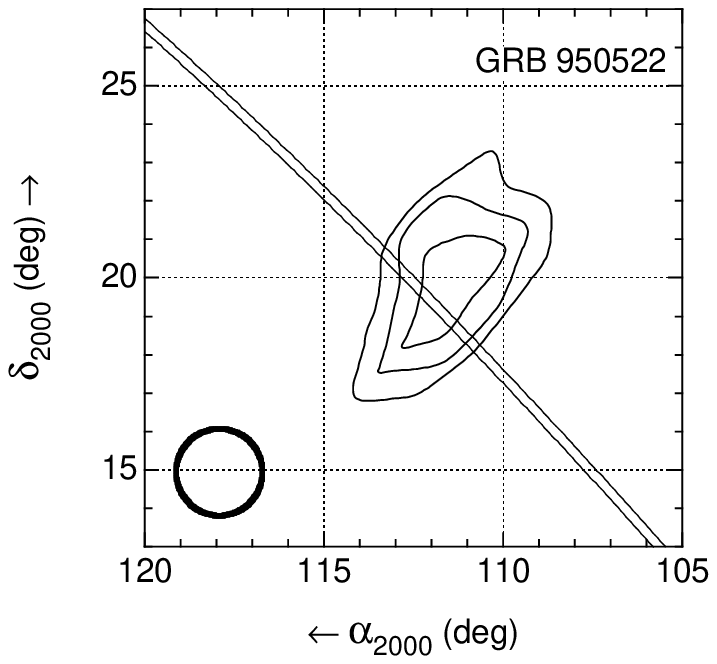}{} 

\medskip
\centerline{\parbox[]{14cm}{
\figcaption[f4.eps]{\baselineskip=9pt\footnotesize Continued.} } }
\end{figure}

\subsection{Measured Systematic Burst Location Error}

As indicated in the sky-maps of Figure~4, 26 of the 29 COMPTEL bursts
have been independently localized through the method of arrival-time
analysis using the CGRO and {\it Ulysses\/} spacecraft as two nodes of
an {\it Interplanetary Network\/} (IPN) baseline (Hurley et al.\
1992).  This technique constrains the direction of a burst to a thin
annulus on the sky, centered at coordinates ($l_{\rm arc}$, $b_{\rm
arc}$) with typical angular radius $R_{\rm arc} \sim 50\arcdeg$ and
width $\sigma_{\rm r} \lesssim 0.1\arcdeg$ (Hurley et al.\ 1996).  The
perpendicular offset $\rho$ between the most probable COMPTEL burst
location ($l$, $b$) and the corresponding IPN annulus is given by

\begin{equation}
  \label{eqrho} 
  \rho = {\rm cos}^{-1} \left [ \sin{b} \sin {b_{\rm arc}} + \cos{b}
  \cos{b_{\rm arc}} \cos{(l-l_{\rm arc})} \right ] - R_{\rm arc}\ ,
\end{equation}

\noindent where positive and negative values indicate whether the
COMPTEL location is inside or outside the annulus, respectively.  This
quantity provides a convenient measure of the total COMPTEL location
error (statistical and systematic) in one dimension.  We have used the
CGRO/{\it Ulysses\/} IPN annuli of 26 bursts to directly estimate the
average COMPTEL systematic burst location error (cf. Graziani \& Lamb
1996).

We began by simulating many COMPTEL ``catalogs'' of burst positions,
with each position randomly blurred according to the statistical error
distributions described by the MLR maps.  An additional systematic
blurring error was applied to the positions in the form of a symmetric
Gaussian probability function with 68\% confidence radius $\sigma_{\rm
sys}$.  For each random burst position, the quantity $\rho_{\rm sim}$
was determined from equation (\ref{eqrho}), assuming that the
corresponding IPN annulus passes directly through the centroid of the
input (for the simulation) burst position.  Uncertainty in the IPN
annulus width was incorporated by assuming a Gaussian error
distribution with standard deviation $\sigma_{\rm r}$.  Using the
Kolmogorov-Smirnov test, we computed the maximum deviation of the
cumulative distribution of observed $\rho$ from that obtained with
2000 blurred catalogs ($\rho_{\rm sim}$).  By varying $\sigma_{\rm
sys}$, and repeating the procedure, we determined how large
$\sigma_{\rm sys}$ must be in order to minimize the deviation between
the observed and blurred distributions.  We found that $\sigma_{\rm
sys} = 0.6\arcdeg \pm 0.4\arcdeg$ yields the best-fit within a 68\%
confidence interval.  The observed and best-fit differential
distributions of $\rho$ are shown in Figure~5.  This measurement is
entirely consistent with the value of $\sigma_{\rm sys} = 0.5\arcdeg$
discussed above, and estimated by independent means.  For all
subsequent analysis we took $\sigma_{\rm sys} = 0.5\arcdeg$ and
assumed that it was a constant for all bursts.  When this error is
added in quadrature to the statistical error radius ($\sigma_{\rm
eff}$), the average COMPTEL location uncertainty increases to
1.25$\arcdeg$.

\begin{figure}[ht]
\centerline{\epsscale{0.5}\plotone{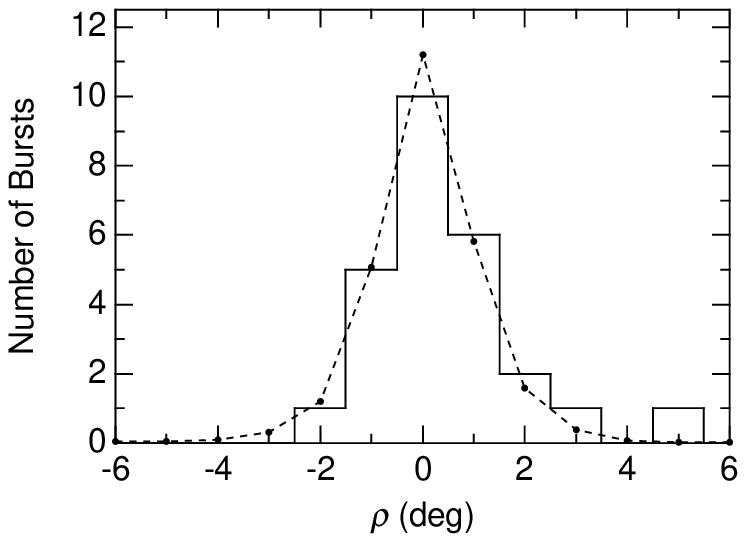}}
\centerline{\parbox[]{14cm}{
\figcaption[f5.eps]{\baselineskip=9pt\footnotesize Distribution of the
angle $\rho$ between most-likely COMPTEL burst positions and IPN
annuli (at closest approach) for 26 bursts.  The solid histogram shows
the measured distribution, while the dashed curve is the best fit
model incorporating statistical COMPTEL and IPN errors and a constant
COMPTEL systematic location error of 0.5$\arcdeg$. \label{fig5} } } }
\end{figure}

\subsection{Combined COMPTEL/IPN Localizations\label{ipnsec}}

The COMPTEL burst localizations are reduced over an order-of-magnitude
in area by incorporating the constraints provided by IPN measurements.
The COMPTEL localizations were used to limit the probable extent of the
IPN annuli in the following manner.  First, the probability
distributions obtained from the observed MLR maps were convolved with a
Gaussian systematic blurring function with $\sigma_{\rm sys} =
0.5\arcdeg$.  This provided, for each burst, a measure of the total
COMPTEL location error distribution about the most-likely position.
The convolved maps were then evaluated at 0.1$\arcdeg$ intervals along
the corresponding IPN annuli---yielding probability as a function of
angular distance along the annuli.  The most probable combined
COMPTEL/IPN location of a burst corresponds to the point of maximum
probability along the arc and confidence intervals were obtained from
the integral probability distribution.  Uncertainty in the width of
the combined locations was assumed to be dominated by IPN measurement
errors that are small compared to COMPTEL errors.

\begin{table}[p]
\epsscale{0.7}
\plotone{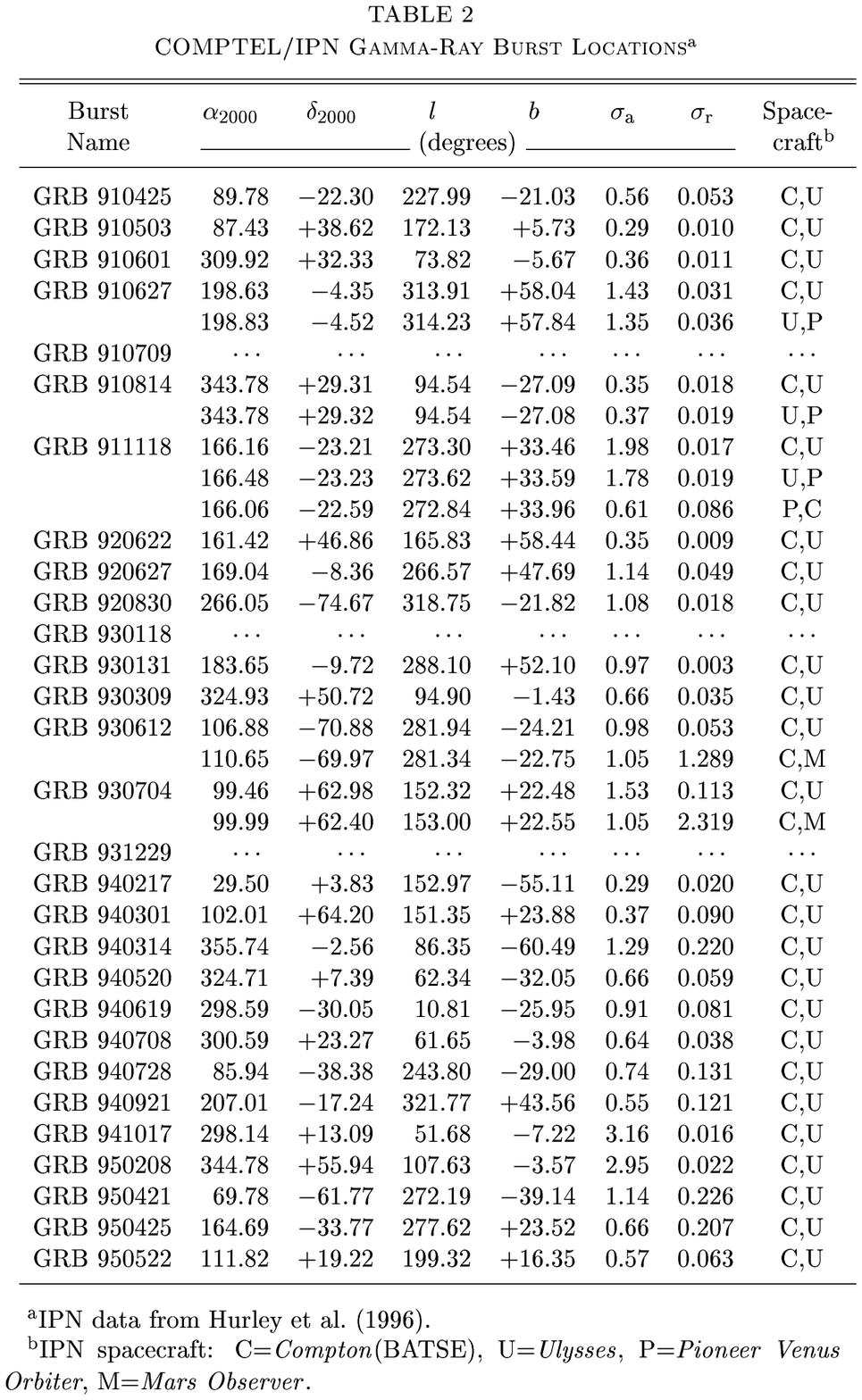}
\label{tab2}
\end{table}

Combined COMPTEL/IPN burst localizations are listed in Table~2, where
the most probable combined locations are expressed in Celestial (epoch
J2000.0) and Galactic coordinates.  The next column gives the mean
68\% confidence error $\sigma_{\rm a}$ in the angular distance along
the arc, as measured from the most probable position.  This is
followed by the uncertainty in the arc width $\sigma_{\rm r}$, as
determined from IPN timing measurements (Hurley et al.\ 1996).  Those
bursts that were observed by more than two spacecraft are indicated.
The average COMPTEL/IPN burst localization (1$\sigma$ confidence)
corresponds to a short, thin arc-segment with total length $2 \times
\langle \sigma_{\rm a} \rangle = 1.97\arcdeg$ and total width $2
\times \langle \sigma_{\rm r} \rangle = 0.13\arcdeg$.  For one burst
(GRB~911118), IPN measurements from three spacecraft provide a nearly
unambiguous determination of the source location where the individual
annuli intersect.  The IPN location of this event is within the
COMPTEL statistical 1$\sigma$ MLR contour (see Figure~4).

\section{Spatial Distribution Analysis}

Despite their limited number, the COMPTEL GRB catalog locations
provide a useful measure of the global angular distribution of GRBs.
In principle, the catalog can also be used to probe the GRB distance
scale through the distribution of burst intensities (e.g., Pendleton
et al.\ 1996; Fenimore et al.\ 1993).  However, the COMPTEL
measurements are ill-suited for this purpose because of severe
observational biases---most notably the strong dependence of burst
sensitivity on fluence (and duration), the large variation of
sensitivity across the FoV and the inability to measure intense burst
episodes due to deadtime.  These biases would require large, uncertain
correction factors at nearly all observed intensity levels.  The
COMPTEL locations can, however, provide some insight into the burst
distance dilemma through correlations with known objects and by
providing a lower limit on the distance to individual events through
comparison with IPN measurements.

We analyzed the distribution of COMPTEL burst locations using two
inclusive samples chosen for easy comparison with the results of
earlier analyses.  The {\it three-year\/} sample includes 18 bursts
observed between 1991 April 19 and 1994 March 1, while the {\it
four-year\/} sample incorporates the full catalog of 29 bursts.
Galactic coordinate distributions of the COMPTEL statistical location
constraints for both burst samples are shown in Figure~6.  The null
hypothesis is that GRB sources have an isotropic angular distribution
on all size scales, with no preferred direction or clustering.  We use
the COMPTEL bursts as a sample of the full GRB population to test this
hypothesis.  The small number of bursts means that statistical
fluctuations are the dominant source of error and that this error is
non-Gaussian.  Statistical tests are therefore conducted by Monte
Carlo methods.  The major observational biases present are the
non-ideal burst locations and uneven sampling of the sky, both of
which affect the interpretation of the data.

\begin{figure}[ht]
\centerline{\epsscale{1.0}\plottwo{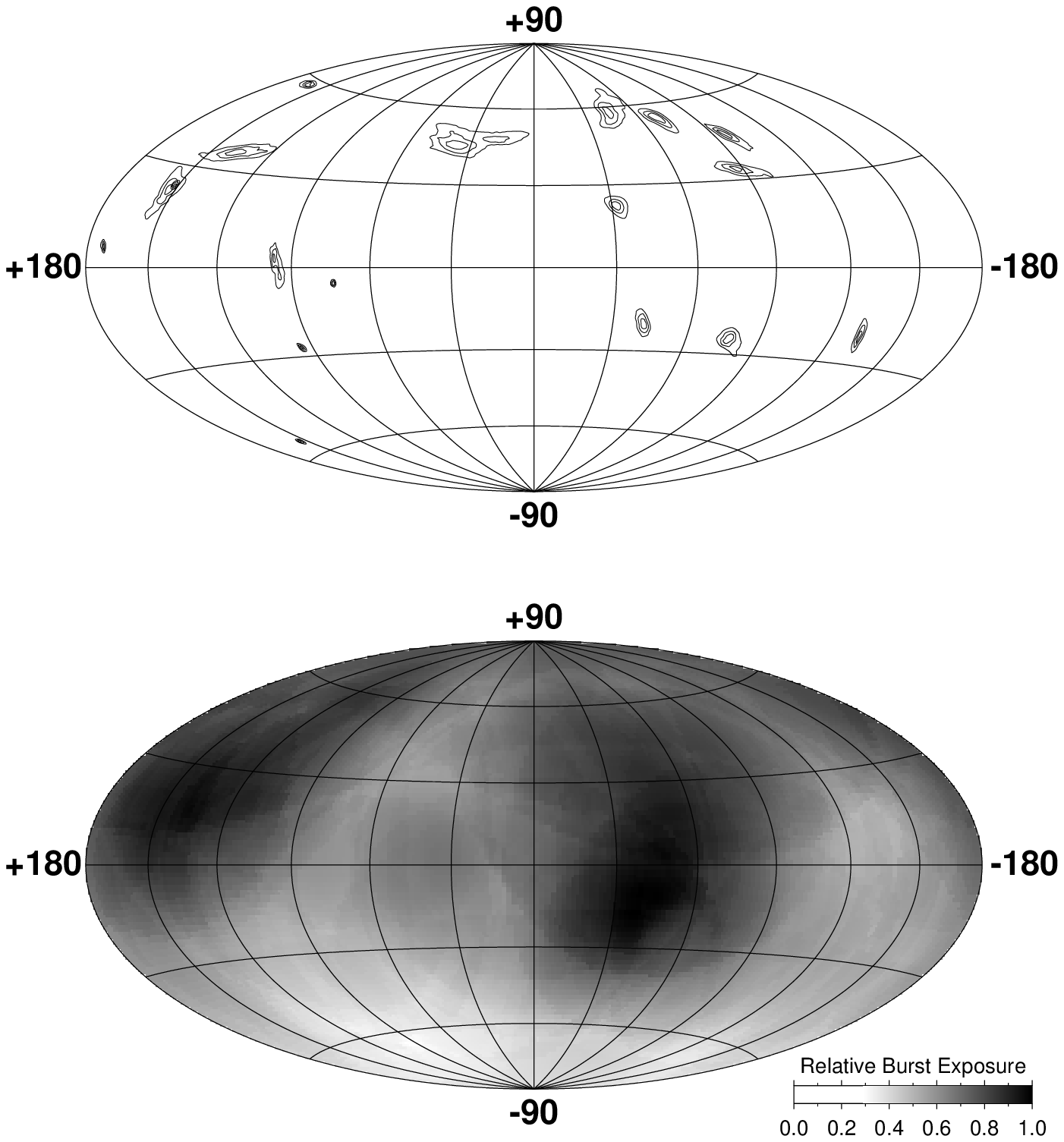}{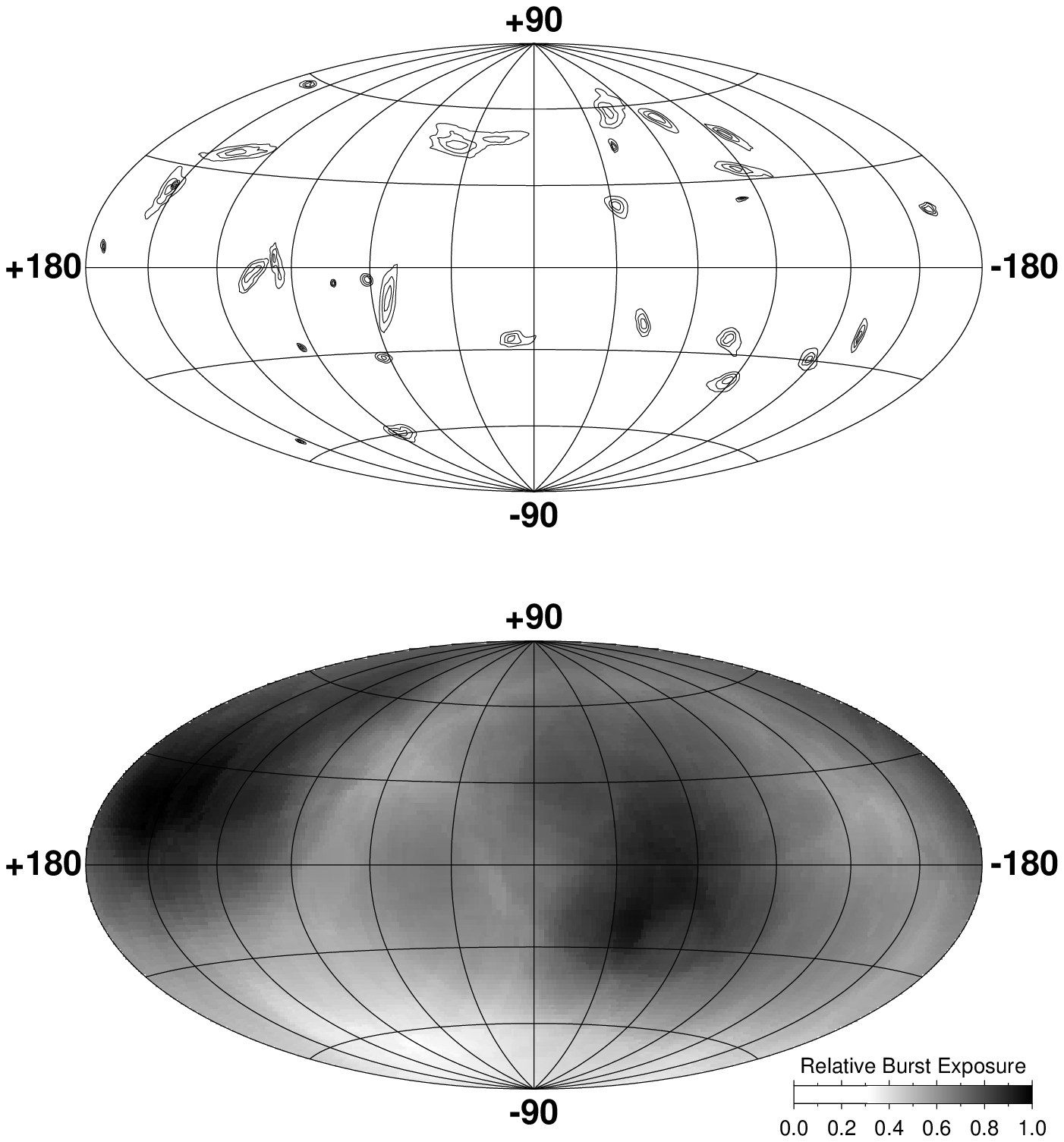}}
\centerline{\parbox[]{14cm}{
\figcaption[f6.eps]{\baselineskip=9pt\footnotesize The COMPTEL
statistical (1$\sigma$, 2$\sigma$ and 3$\sigma$) location contours
({\it top}) and corresponding burst exposure functions ({\it bottom})
plotted in Galactic coordinates.  The leftmost plots correspond to the
three-year observation period (18 bursts), while the plots at right
include the full four-year interval (29 bursts). \label{fig6} } } }
\end{figure}

\subsection{Burst Exposure Function\label{expsec}}

The limited extent and varying sensitivity of the COMPTEL FoV,
combined with CGRO's low Earth orbit and sporadic pointing schedule
result in a complicated sampling pattern over the sky that modifies
the intrinsic angular source distribution.  We must determine this
pattern---the {\it burst exposure function\/}---in order to correct
the observational bias it produces.

Assuming that bursts occur isotropically in space and uniformly in
time, the relative probability $P_{\rm exp}$ of detecting a burst from
a particular direction in the sky compared to any other direction is
given by

\begin{equation}
   \label{eqlpexp}
   P_{\rm exp} = t_{\rm on} \times \int_{S_{\rm min}}^{\infty} 
   \frac{dN}{dS}\,dS\ ,
\end{equation}

\noindent where $t_{\rm on}$ is the time spent observing the direction
of interest, $S_{\rm min}$ is the minimum fluence for a detection and
$\case{dN}{dS}$ is the differential burst size--frequency distribution
(i.e., the number of bursts per year with 0.75--30 MeV fluence between
$S$ and $S + dS$).  Evaluation of $P_{\rm exp}$ is complicated by the
function $S_{\rm min}$, that depends on the source position within the
FoV, the background rate and the burst energy spectrum.  We
approximated $P_{\rm exp}$ by ignoring background and taking the burst
energy spectrum to be the average power law ($E^{-2.6}$; Kippen et
al.\ 1996a, 1997)---leaving $S_{\rm min}$ as a function only of
sky-position through equation (\ref{eqsmin}).  A further problem is
that $\case{dN}{dS}$ is not well-known in the COMPTEL energy range.
For simplicity, we used a power law of the form $\case{dN}{dS} \propto
S^{-3/2}$.  For strong bursts, such as those observed by COMPTEL, this
is a reasonable approximation (Fenimore et al.\ 1993).

The above approximations were used to compute $P_{\rm exp}$ for every
16 seconds of observation on a grid of sky coordinates with $2\arcdeg
\times 2\arcdeg$ spacing.  Directions occulted by the earth were
excluded, as were all epochs when COMPTEL and/or the BATSE trigger
system were disabled or the spacecraft was not telemetering data.
Directions outside the $\vartheta_{\rm c} \leq 65\arcdeg$ search
window were also excluded.  Sky-maps of the burst exposure function
accumulated over the two analysis intervals are shown in Figure~6.
The maximum value of $P_{\rm exp}$ in the three-year (four-year)
exposure map is 426 (624) effective days.  In both maps, the minimum
exposure is at least 30\% of the maximum---meaning that all portions
of sky have been observed at a reasonable level.  The smooth
fluctuations on top of this minimum exposure reflect the frequent
Earth occultation of directions along the Celestial equator and the
schedule of the CGRO pointing program.  In particular, the south
Galactic polar region received the least exposure, while a great deal
of time was spent observing the Galactic plane and Virgo
($l=270\arcdeg$, $b=60\arcdeg$) regions.  The effects of different
spectral shapes, burst intensity function and the inclusion or
exclusion of BATSE data accumulation periods are small
($\lesssim$10\%) compared to the overall magnitude of the structures
in the exposure maps.

\subsection{Large-Scale Angular Distribution}

Studies of large-scale angular structure have been most concerned with
the question of whether burst sources reside inside or outside our
Galaxy.  Two convenient and sensitive statistical tools for this
particular purpose are the dipole and quadrupole moments in Galactic
coordinates (e.g., Hartmann \& Epstein 1989; Paczy\'{n}ski 1990;
Briggs 1993).  The dipole moment (expressed as $\langle \cos{\Theta}
\rangle$, where $\Theta$ is the angle between a burst and the Galactic
center) measures the amount of concentration towards the center of the
Galaxy, while the quadrupole moment (expressed as $\langle \sin^2{b} -
\case{1}{3} \rangle$, where $b$ is the Galactic latitude of a burst)
indicates the amount of concentration towards the Galactic plane.
Significant non-zero values of the moments indicate deviations from an
isotropic source distribution.  Similar statistics, expressed in
Celestial coordinates (i.e., $\langle \cos{\Theta} \rangle$, where
$\Theta$ is the angle from the Celestial origin and $\langle
\sin^2{\delta} - \case{1}{3} \rangle$, where $\delta$ is declination),
are more sensitive to the biases caused by the non-uniform COMPTEL
exposure function.

The dipole and quadrupole moments expected for an isotropic angular
distribution of sources are modified by the non-uniform burst exposure
function.  The magnitude of this effect was estimated by simulating
random catalogs of burst locations chosen according to the COMPTEL
exposure function maps.  An ensemble of moments computed from $5
\times 10^4$ simulated catalogs provides a measure of the
exposure-induced deviation from isotropy and the statistical
uncertainty inherent in the number of bursts in each catalog.  The
largest deviation from the isotropic value appears in the Celestial
quadrupole moment, where the exposure function produces a 0.7$\sigma$
bias towards polar declinations (both three and four-year
samples). The exposure-induced bias in Galactic coordinates is much
smaller (see examples shown in Figure~7).

\begin{figure}[ht]
\centerline{\epsscale{0.5}\plotone{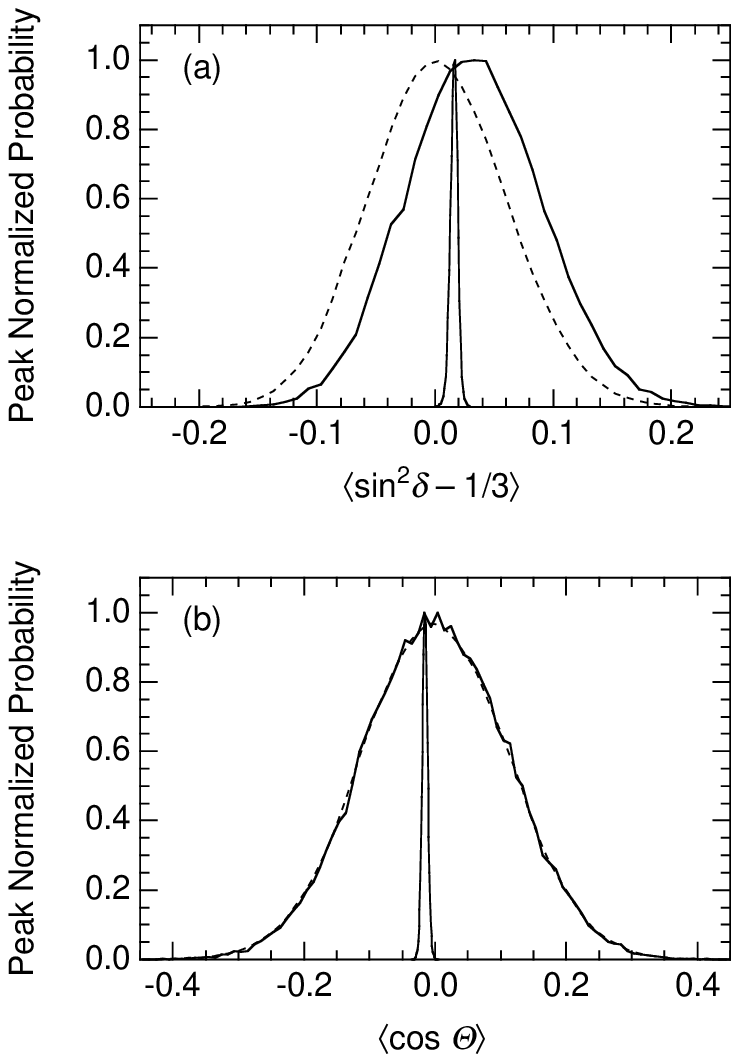}}
\centerline{\parbox[]{14cm}{
\figcaption[f7.eps]{\baselineskip=9pt\footnotesize Simulations of
Celestial quadrupole moment (a) and Galactic dipole moment (b)
distributions for 29 burst locations.  The narrowest distributions in
each case represent the values observed by COMPTEL, including the
effect of location errors.  The remaining curves show the statistical
distributions expected from isotropy assuming uniform ({\it dashed
curves\/}) and non-uniform ({\it solid curves\/}) burst exposure
functions. \label{fig7} } } }
\end{figure}

Observed moments were estimated by randomly blurring each of the
measured burst locations according to the total (statistical and
systematic) location error distributions.  An ensemble of moments
computed from many blurred catalogs provides a measure of the error
induced by imprecise locations.  As illustrated in Figure~7, this
effect is small compared to the statistical uncertainty caused by the
low number of bursts.  The difference between the observed and
expected moments yields the true or corrected deviation from isotropy.
Corrected values of the dipole and quadrupole moments are given in
Table~3 in both Galactic and Celestial coordinates.  Uncertainties in
the corrected moments reflect 68\% confidence limits on the difference
between the data and the simulations.  Within the sizable
uncertainties, there are no significant deviations from large-scale
isotropy evident in the COMPTEL burst samples in either Galactic or
Celestial coordinates.

\begin{table}[ht]
\epsscale{0.8}
\plotone{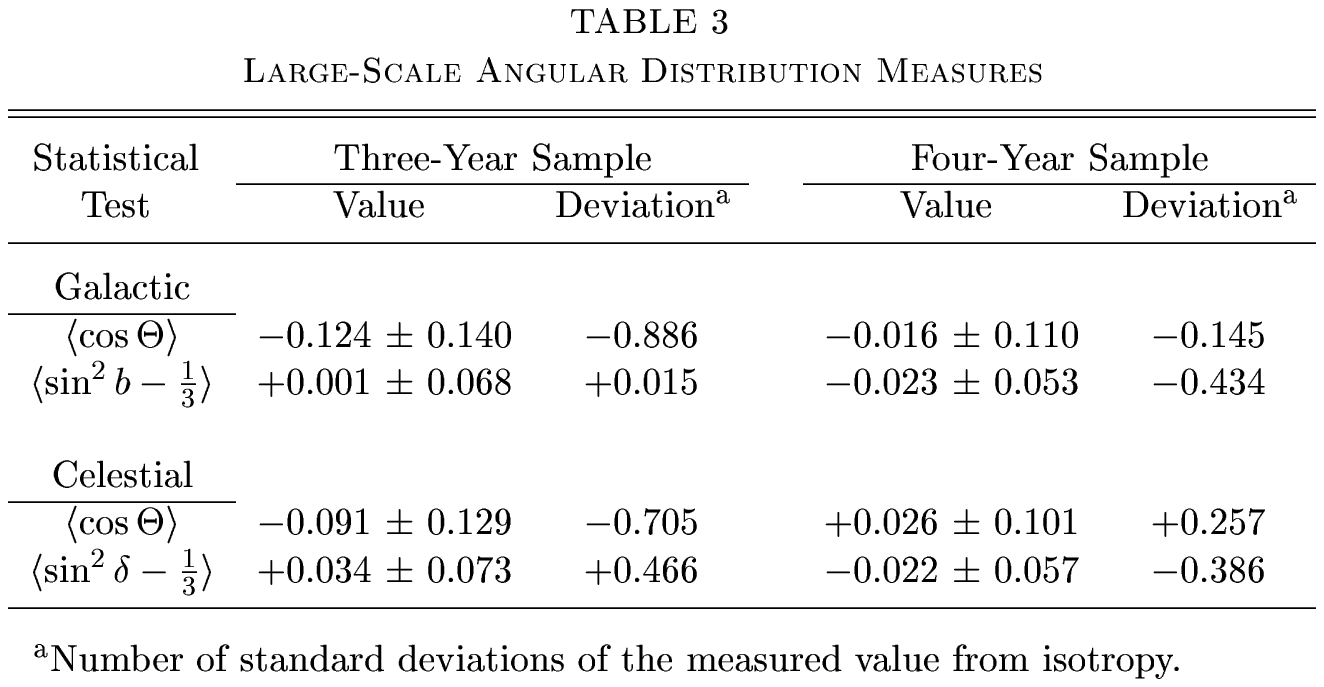}
\label{tab3}
\end{table}

\subsection{Small-Scale Angular Distribution}

The dipole and quadrupole moments are sensitive only on the largest of
angular scales ($\theta \gtrsim 60\arcdeg$).  Different statistical
tests are necessary to investigate the small-scale angular distribution
of burst sources.  The two standard approaches are the angular
autocorrelation function $w(\theta)$ and the nearest neighbor function
$n\!n(\theta)$.  Both statistics are sensitive to small-scale angular
structure on the order of the instrumental resolution, which could
suggest either intrinsic source clustering or repeated bursts from
individual sources.  It has been shown that the angular correlation
function provides superior sensitivity to small-scale angular
structure when the number of sources is large, such that the average
separation between neighboring positions is less than the average
location accuracy (Brainerd 1996).  This is not the case for the small
sample of well-localized COMPTEL bursts, where the average neighbor
separation ($\sim$20$\arcdeg$) is much greater than the average
location accuracy ($\sim$1.25$\arcdeg$).  Hence, both tests were used.

The angular correlation function, first applied to GRB positions by
Hartmann \& Blumenthal (1989) and Hartmann, Linder \& Blumenthal
(1991), measures the probability of finding a pair of sources
separated by an angle $\theta$.  It is computed by comparing the
number of source pairs $N_{\rm obs}$, with angular separations in the
range $(\theta, \theta + \rm{d}\theta)$, to the number of pairs
$N_{\rm exp}$ expected from a random sample of an isotropic
distribution with 

\begin{equation}
   \label{eqacor}
   w(\theta) = {N_{\rm obs} \over N_{\rm exp}}  - 1\ .  
\end{equation}

\noindent The presence of small-scale angular clustering is evident in
$w(\theta)$ as a significant positive deviation from zero at small
angles.  The nearest neighbor function $n\!n(\theta)$, first used on
GRB positions by Quashnock \& Lamb (1993), differs slightly in that it
compares the number of sources with one or more neighbors separated by
$(\theta, \theta + \rm{d}\theta)$ to the number expected in a random,
isotropic sample.  Small-scale clustering is evident in $n\!n(\theta)$
as a significant excess above the expected value at small angles.
Both statistics are affected by imprecise source locations, which
spread the intrinsic signal of clustering over a range of $\theta$
characteristic of the average location accuracy.  The signal of
small-scale clustering is maximized by choosing bins in $\theta$ with
the same characteristic size.

Figure~8 shows the angular correlation and nearest neighbor
distributions for the two samples of COMPTEL burst locations.  Angular
bins of 2$\arcdeg$ were used to approximate the average location
accuracy at the 90\% confidence level, where the signal from
clustering would be most apparent.  The distributions of $N_{\rm exp}$
expected from isotropy were determined by taking the average of
$5 \times 10^4$ simulated catalogs, each chosen at random according to
the corresponding burst exposure function.  The effect of non-uniform
exposure on the $N_{\rm exp}$ distributions is small compared to
statistical fluctuations among the random catalogs ($< 0.1\sigma$).
The statistical significance of observed deviations in excess of the
isotropic expectation was estimated by the fraction of simulated
catalogs $Q$ with an equal or greater excess in the angular bin in
question.

\begin{figure}[ht]
\centerline{\epsscale{1.0}\plotone{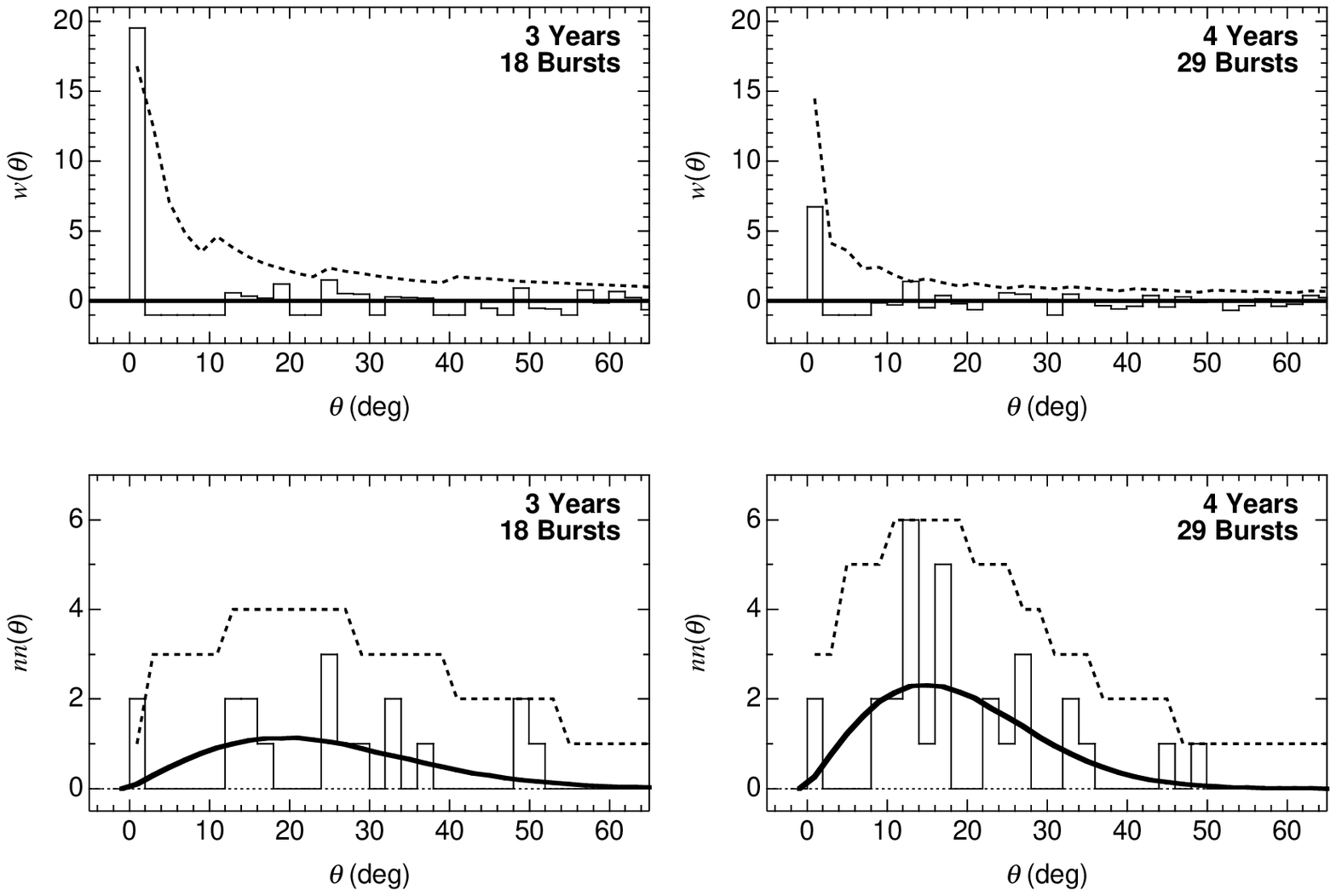}}
\centerline{\parbox[]{14cm}{
\figcaption[f8.eps]{\baselineskip=9pt\footnotesize Angular
auto-correlation functions $w(\theta)$ and nearest neighbor
distributions $n\!n(\theta)$ of two samples of COMPTEL burst
locations.  Each plot compares the distribution measured with the
most-likely locations ({\it solid histograms\/}) to the distributions
expected from the non-uniform sampling of isotropy ({\it solid
curves\/}).  Dashed curves indicate the magnitude of statistical
deviations from the expected distributions at the 90\% significance
level. \label{fig8} } } }
\end{figure}

The most significant small-angle excess is evident in the first
2$\arcdeg$ bin in the three-year catalog distributions, where $Q = 4.8
\times 10^{-2}$.  This excess occurs because two of the bursts
(GRB~930704 and GRB~940301) are localized to the same direction within
their combined location errors---something that is not expected in a
random sample of only 18 sources (Ryan et al.\ 1994a).  This near
coincidence of two bursts occurring eight months apart from one
another was investigated by Kippen et al.\ (1995b) and Hanlon et al.\
(1995), who computed a similar chance probability for the observation
($Q = 3.0 \times 10^{-2})$.  The value of $Q$ computed here is
somewhat larger due mainly to the addition of the previously neglected
burst GRB 920627.  Kippen et al.\ (1995b) could neither prove, nor
rule out the possibility that the two bursts originated from a single
source---even when all other available measurements were included
(IPN, BATSE and CGRO--EGRET).  In the larger four-year sample the
significance of the excess at $\theta < 2\arcdeg$ is reduced to $Q =
1.2 \times 10^{-1}$---a result of the fact that no additional bursts
from the possible repeating source were observed in the year following
GRB~940301, and that no other coincident pairs were measured
elsewhere.  We therefore conclude that the four-year sample is
consistent with isotropy on small scales.

\subsection{Correlation with Galaxy Catalogs}

Examination of correlations between GRBs and catalogs of known objects
is similar to the study of self-clustering described above.
Statistical tools like the nearest neighbor test and the angular
correlation function are simply modified by computing angles between
GRB/object pairs, rather than angles between pairs of different
bursts.  As in the case of self-clustering, cross correlations between
GRBs and known objects are affected by burst location errors, which
spread the cross-correlation signal over an angular range
characteristic of the average burst location error.  Although many
different types of astronomical catalogs have been examined, we chose
to concentrate only on galaxy clusters and radio-quiet quasars, which
are suggested to correlate with BATSE 3B burst locations (Kolatt \&
Piran 1996; Marani et al.\ 1997; Schartel, Andernach \& Greiner 1997).
In particular, we used the galaxy cluster catalog of Abell, Corwin \&
Olowin (1989, hereafter ACO), and the quasar catalog of
V\'{e}ron-Cetty \& V\'{e}ron (1996, hereafter VCV).  The full ACO
catalog (ACO-1) contains 5250 galaxy clusters, categorized by several
parameters, including richness class $R$ and distance class $D$.  A
subset of the 185 ACO galaxy clusters with $R \geq 1$ and $D \leq 4$
(ACO-2) was examined separately since this particular subset was
suggested to show a correlation with the most accurate BATSE burst
locations (Marani et al.\ 1997 and references therein).  The VCV
catalog contains $11,662$ quasars and active galactic nuclei.  We
examined only the radio-quiet quasars, as selected by Schartel,
Andernach \& Greiner (1997).  The full radio-quiet quasar sample
(VCV-1) contains 7146 objects.  A subset of 967 radio-quiet quasars
with redshift $z \leq 1.0$ and absolute magnitude $M_{\rm abs} \leq
-24.2$ (VCV-2) was also examined since this sample shows the strongest
correlation with BATSE GRBs.

\begin{figure}[p]
\centerline{\epsscale{0.5}\plotone{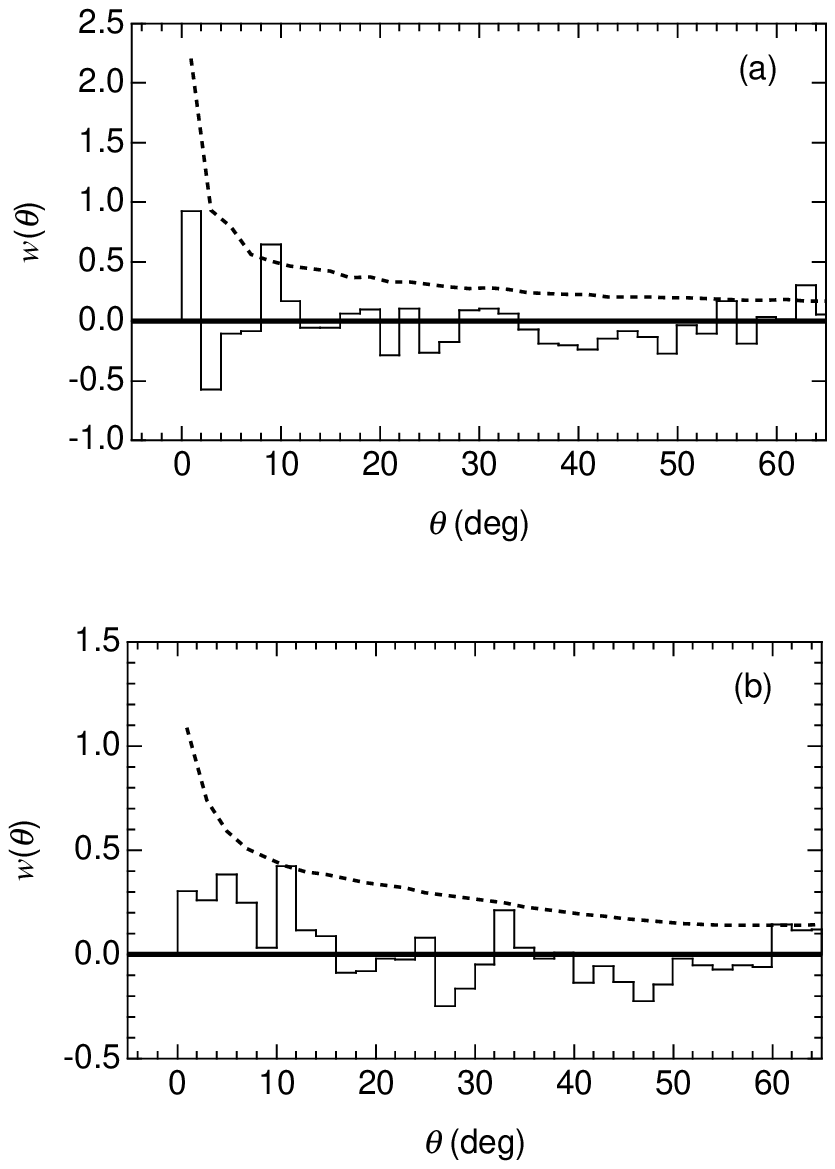}}
\centerline{\parbox[]{14cm}{
\figcaption[f9.eps]{\baselineskip=9pt\footnotesize Angular
cross-correlation function $w(\theta)$ between 29 COMPTEL burst
locations and: (a) 185 ACO galaxy clusters with richness class $R \geq
1$ and distance class $D \leq 4$; (b) 967 VCV radio-quiet quasars with
redshift $z \leq 1.0$ and absolute magnitude $M_{\rm abs} \leq -24.2$.
Each plot compares the distribution observed with the most-likely
COMPTEL burst locations ({\it solid histograms\/}) to the distribution
expected, on average, from random samples of the non-uniform COMPTEL
exposure function ({\it solid lines\/}).  Dashed curves indicate the
magnitude of statistical deviations from the expected distributions at
the 90\% significance level. \label{fig9} } } }
\end{figure}

\begin{table}[p]
\epsscale{0.5}
\plotone{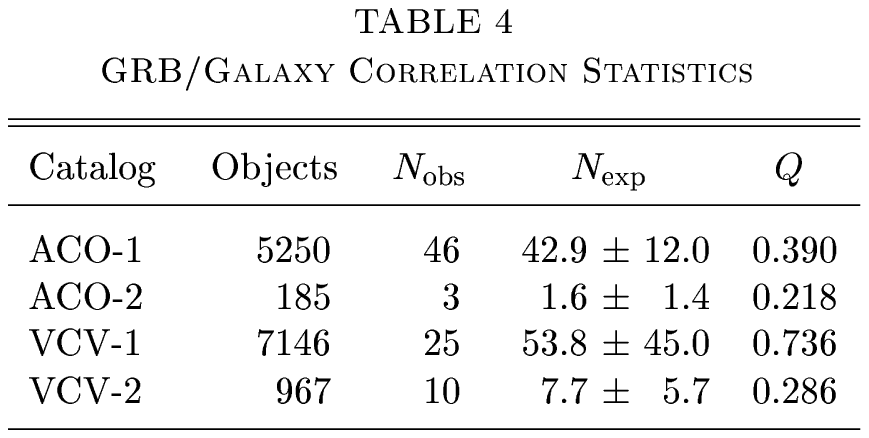}
\label{tab4}
\end{table}

The computation of angular cross-correlation between COMPTEL GRBs and
ACO/VCV catalog locations was performed as in the case of
self-clustering.  In each 2$\arcdeg$ angular bin the number of
observed GRB/catalog pairs $N_{\rm obs}$ was compared to the average
number obtained from $5 \times 10^4$ random COMPTEL ``catalogs''
$N_{\rm exp}$ (sampled from the four-year COMPTEL burst exposure
function) through equation (\ref{eqacor}).  Table~4 lists the results
for GRB/catalog pair angular separations $\theta < 2\arcdeg$.  The
uncertainty in $N_{\rm exp}$ given in the table reflects the standard
deviation of the random catalogs and $Q$ is the statistical
significance of the observed number of pairs compared to the expected
number.  None of the four catalog samples show a significant
correlation with COMPTEL GRB locations at small angles, where one
expects the signal if an association exists.  Angular
cross-correlation functions for the ACO-2 and VCV-2 object catalogs
are shown in Figure~9.

\subsection{COMPTEL/IPN Burst Distance Limits}

While a ``local'' GRB source population has not been examined in as
much detail as the more distant possibilities, such an origin (i.e.,
sources within $10^4$ AU) is not inconsistent with the angular and
intensity distributions observed by BATSE (Maoz 1993; Horack et al.\
1994).  Lower limits on the distance to burst sources are therefore
important.  As pointed out by Hurley (1982), timing differences
between four or more widely separated IPN spacecraft in a favorable
configuration could be used to yield a lower limit on the burst source
distance.  While only a few events have been observed by this many IPN
spacecraft, the idea can also be applied to the bursts localized by
COMPTEL and the two-spacecraft, CGRO/{\it Ulysses\/} IPN.

If a burst is observed by two spacecraft separated by a distance $b$,
the angle $\alpha$ between the burst direction and the
inter-spacecraft baseline is

\begin{equation}
   \label{eqdipn}
   \cos{\alpha} = \left(\frac{c\Delta t}{b}\right) + \frac{b}{2D} 
   \left\{1 -  {\left(\frac{c\Delta t}{b}\right)}^2 \right\}\ , 
\end{equation}

\noindent where $c$ is the speed of light, $D$ is the distance to the
burst source and $\Delta t$ is the observed difference in burst
arrival times between the two spacecraft (Connors et al.\ 1993a).
Thus, $\cos{\alpha}$ is larger than that inferred from the time-delay
alone ($\cos{\alpha} = \cos{R_{\rm arc}} = c\Delta t / b$) by a
parallax factor that is inversely related to the source distance.  If
the offset of a burst position from an IPN annulus were measured, it
would reveal the source distance.  The fact that COMPTEL burst
localizations agree with IPN annuli to within the COMPTEL location
accuracy places a lower limit on the distance to burst sources.  The
severity of this limit improves with the accuracy of the COMPTEL
localization (in the direction perpendicular to the IPN annulus) and
the length of the IPN baseline ($b$).

\begin{table}[ht]
\epsscale{0.5}
\plotone{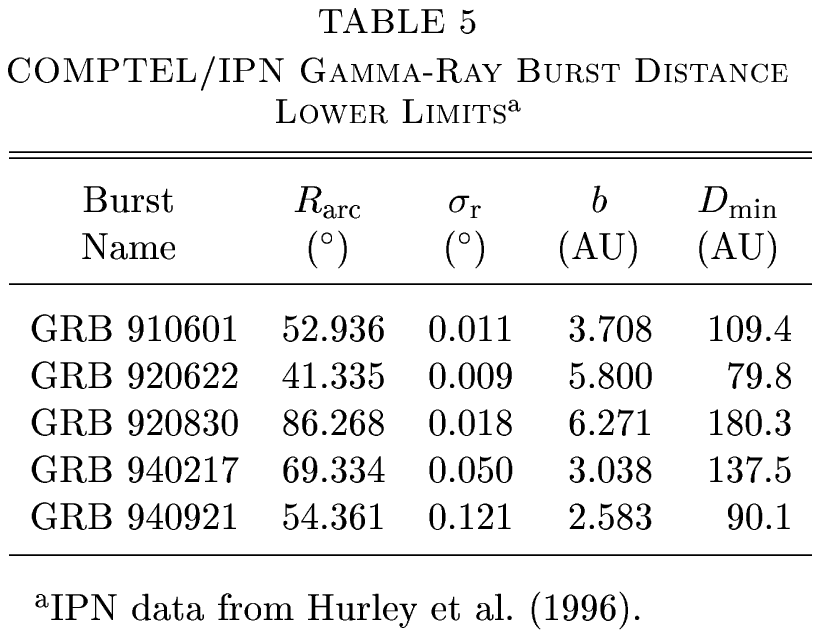}
\label{tab5}
\end{table}

According to the procedure described in \S\ref{ipnsec}, the total
(statistical and systematic) COMPTEL localization of a burst can be
maximized along an annulus to reveal the most-likely combined position
and the probability at that position.  In computing a quantitative
COMPTEL/IPN distance limit this technique was applied to a set of
concentric annuli with angular radii $\alpha$ progressively smaller
than $R_{\rm arc}$.  This yielded---with the use of equation
(\ref{eqdipn})---the COMPTEL location probability as a function of
source distance $D$.  Uncertainty in the width of the IPN annuli was
incorporated by substituting $R_{\rm arc} + \sigma_{\rm r}$ in the
calculation of probability as a function of $D$.  Integration of this
function at the desired confidence level yields a lower limit.  The
most constraining distance lower limits ($D_{\rm min}$) at the
2$\sigma$ confidence level are listed in Table~5.  Even when all known
systematic location errors are conservatively included, the distance
to these bursts must be at least $\sim$100 AU in order to be
consistent with the COMPTEL/IPN measurements.

\section{Summary and Discussion}

We have demonstrated the ability of COMPTEL to accurately localize GRB
sources through direct Compton scatter imaging.  This ability has been
used to accumulate a unique collection of locations for bursts having
a high fluence of gamma rays in the energy range 0.75--30 MeV.
Analysis of the COMPTEL locations has shown that these bursts,
apparently like all others, are consistent with an isotropic angular
distribution of sources.  Furthermore, we find no statistically
significant excess of GRB locations clustered at small angles, which
would be expected if many of the sources we observed burst repeatedly
on time-scales of several months.  We also find no significant
cross-correlation with ACO galaxy clusters or radio-quiet VCV quasars,
and have placed a lower limit on the distance to GRB sources by
comparing the COMPTEL locations to IPN timing annuli.  The COMPTEL
measurements thus contribute to our understanding of the GRB
phenomenon in each of the three currently feasible source distance
regimes.

For a local heliospheric origin, the distance constraints required by
COMPTEL/IPN burst location measurements dispute any scenarios in which
the sources of brighter (and therefore closer, assuming bursts are
roughly ``standard luminosity candles'') GRBs are nearer than
$\sim$100 AU (approximately the distance of the solar wind termination
shock).  Relativistic dust grain models (Grindlay \& Fazio 1974;
Dasgupta 1979) and nearby (40--400 AU) colliding comet scenarios
(White 1993, 1994) are therefore unfavorable.  More distant Oort cloud
models involving mergers between comets, antimatter and/or primordial
black holes (Katz 1993; Dermer 1996; Bickert \& Greiner 1993) are not
excluded by our distance limits.  However, dynamical simulations of
potential Oort cloud GRB progenitor distributions indicate that it is
difficult to restrict GRBs only to the outer, more distant portions of
the cloud (Clarke, Blaes \& Tremaine 1994).  In order to obey the
angular and intensity distributions measured with BATSE {\it and\/}
the distance lower limits reported here, Oort cloud models must invoke
an {\it ad hoc\/} assumption that the GRB progenitor distance has a
sharp cutoff between 100--500 AU (e.g., Horack et al.\ 1994).  Future
instruments capable of more accurate GRB localization (e.g., BeppoSAX
and HETE-2) may increase the distance limit.  For example, a burst
localized with arc-minute accuracy could (when compared to an IPN
annulus) raise the distance lower limit to thousands of AU.

The Galactic dipole and quadrupole moments of COMPTEL burst locations
independently confirm the long-known result that strong bursts are
distributed isotropically (Atteia et al.\ 1987; Hartmann \& Epstein
1989).  In addition, the COMPTEL measurements mean that GRBs with
predominantly hard spectra are probably from the same angular source
distribution as the main bulk of events observed by BATSE, even though
the radial distribution of soft and hard events measured with BATSE
intensities may differ (e.g., Belli 1995; Kouveliotou et al.\ 1996).
These results are important in the context of Galactic halo scenarios,
where one expects all the nearer, brighter bursts to exhibit a
significant positive dipole moment due to the 8.5 kpc offset of the
Sun from the Galactic center.  Unfortunately, the absence of a dipole
in the COMPTEL bursts is not constraining due to the large
uncertainties involved.

In the cosmological distance regime, the absence of significant
small-angle correlations among the COMPTEL locations is important,
since it bears upon the question of burst repetition.  Given that the
larger four-year sample of bursts does not show any significant
evidence for small-scale clustering, we must conclude that the
two-burst overlap between GRB~930704 and GRB~940301 was either a
statistical fluke, or that the two bursts were from a rare repeating
source (or source cluster), whose rate of repetition at the COMPTEL
sensitivity level is low.  If the latter case is true, the lack of
excess BATSE bursts from the GRB~930704/GRB~940301 region indicates
that the repetition rate from this source is also low at the BATSE
sensitivity level.  The rarity of repeating GRB sources can be
estimated from the COMPTEL observations, where the fraction of
observed repeating burst sources must be less than $2/29 = 7\%$.  We
can obtain this number without the use of complicated statistical
tests, as are required in the analysis of BATSE locations (e.g., see
Meegan et al.\ 1995; Tegmark et al.\ 1996), since any other possible
repeater candidates would be immediately obvious given the small
number of COMPTEL bursts and the accuracy of the locations.  It is
interesting that limits on the fraction of repeating burst sources
obtained with BATSE (incorporating thousands of bursts) are about the
same as the 7\% given here, even though BATSE detects bursts that are
an order of magnitude weaker (Meegan et al.\ 1996; Tegmark et al.\
1996).  It thus appears that either: a) repeating burst sources are
rare, b) the typical repetition time-scale is $\gtrsim$1 yr, or c)
repeated bursts are too weak to be reliably detected, even by BATSE.
More constraining results will come from continued observations and
from future instruments with improved location accuracy.

Of further importance in the cosmological distance regime is the lack
of significant correlation between COMPTEL bursts and either ACO
galaxy clusters or radio-quiet quasars.  These findings question the
reality of BATSE/ACO/VCV associations that have been suggested.
According to Marani et al.\ (1997) the strongest cross-correlation
signal for BATSE/ACO clusters ($\sim$3.5-4$\sigma$ significance) is
obtained for the 27 BATSE 3B bursts with the smallest location errors
and that the signal weakens as bursts with larger location errors are
included.  This could indicate that bright (or high fluence) bursts
are more highly correlated with ACO galaxy clusters than weak events,
since bright bursts tend to yield smaller statistical location errors
than weaker bursts.  This behavior is exactly what we would expect of
cosmological GRBs, where only the closest, brightest events would be
correlated with relatively nearby ACO galaxies.  However, the same
correlation should exist in the COMPTEL bursts, since they are the
same average brightness and fluence as the 27 well-localized BATSE 3B
events.  In fact, the two samples share six bursts in common.  The
only significant difference between the COMPTEL and BATSE burst
samples is that the COMPTEL events have harder spectra, on average, than
the BATSE bursts.  It is possible that hard events are not as strongly
correlated with ACO galaxy clusters as are soft bursts, but it is more
likely that the BATSE/ACO correlation is a statistical fluke.  This
conclusion is supported by Hurley et al.\ (1997), who find that a
majority of the ACO clusters that are correlated with BATSE locations
are inconsistent with IPN timing annuli---implying that the BATSE/ACO
correlation is statistical in origin.  The lack of a correlation
between COMPTEL burst locations and VCV radio-quiet quasars is less
meaningful, since the number of COMPTEL bursts is much smaller than
the number of BATSE events (80 bursts with location errors
$< 1.7\arcdeg$) used by Schartel, Andernach \& Greiner (1997).  The
weak GRB/quasar correlation using the more numerous BATSE locations
may be below the sensitivity of the COMPTEL sample due to its larger
statistical uncertainty.

While the collective analysis of COMPTEL burst locations offers some
insight into the GRB mystery, perhaps the greatest value of the
locations is in facilitating rapid counterpart search efforts.  The
recent discovery of x-ray and optical emission fading on a time-scale
of hours/days from the BeppoSAX locations of GRB 970228 and GRB 970508
(Costa et al.\ 1997a, 1997b; van Paradijs et al.\ 1997; Bond 1997)
highlights the value of rapid follow-up studies using accurate burst
locations.  During the course of the CGRO mission, we have developed a
rapid response system to quickly compute COMPTEL GRB locations and
distribute them to multi-wavelength observers (primarily optical and
radio) around the world (Kippen et al.\ 1993).  The advantage of using
the relatively accurate COMPTEL locations is that they allow deeper
searches than previous, and most other current efforts.  The system
has been used for several bursts, resulting in some of the most
sensitive radio/optical observations of burst locations ever obtained
within several hours (see e.g., Harrison et al.\ 1995; McNamara et
al.\ 1996 and references therein).  While no fading counterparts have
been detected as a result of the COMPTEL rapid response effort, the
observations obtained point the way for future studies, which must
concentrate on deep exposures within minutes of a GRB.  This is the
goal of continuing COMPTEL rapid response efforts, as well as several
on-going and planned new missions.  If the optical lightcurves of GRB
970228 and GRB 970508 are typical then wide-field instruments guided
by a COMPTEL GRB location within $\sim$1 hour may detect the fading
counterpart.

\acknowledgments

We thank C. Meegan and the BATSE operations team for timely
communication of BATSE trigger information and N. Schartel for
providing his selected quasar catalogs.  This research was supported,
in part, through the CGRO guest investigator program under grant
NAG5-2350.  The COMPTEL project is supported by NASA under contract
NAS5-26645, by the Deutsche Agentur f\"ur Raumfahrtangelegenheiten
(DARA) under grant 50 QV90968 and by the Netherlands Organization for
Scientific Research (NWO).  DHH acknowledges NASA support under grant
NAGW5-1578 and KH is grateful for support under JPL Contract 958056 for
{\it Ulysses\/} operations, and NAG5-1560 for the Interplanetary
Network.  This research made use of data obtained electronically from
the High Energy Astrophysics Science Archive Research Center (HEASARC)
and from the Astronomical Data Center (ADC).


\begin{references}

\reference{} Abell, G. O., Corwin, H. G., \& Olowin, R. P. 1989, \apjs, 70, 1

\reference{} Atteia, J.-L., et al. 1987, \apjs, 64, 305

\reference{} Belli, B. M. 1995, \apss, 231, 43

\reference{} Bennett, K., et al.\ 1993, IAUC, 5749

\reference{} Bickert, K. F., \& Greiner, J. 1993, in AIP
   Conf. Proc. 280, Compton Gamma Ray Observatory, ed. M. Friedlander,
   N. Gehrels, \& D. J. Macomb (New York: AIP), 1059

\reference{} de Boer et al.\ 1991, in Data Analysis in Astronomy IV,
   ed.  V. Di Ges\`u, et al.\ (New York: Plenum Press), 241

\reference{} Bond, H. E. 1997, IAUC, 6654

\reference{} Brainerd, J. J., et al.\ 1995, \apj, 441, L39

\reference{} Brainerd, J. J. 1996, \apj, 473, 974

\reference{} Briggs, M. S. 1993, \apj, 407, 126

\reference{} Briggs, M. S., et al.\ 1996, \apj, 459, 40

\reference{} Connors, A., et al.\ 1993a, \aaps, 97, 75

\reference{} Connors, A., et al.\ 1993b, Adv. Space Res., 13, No. 12, 715

\reference{} Costa, E., et al.\ 1997b, IAUC, 6649

\reference{} Costa, E., et al.\ 1997a, IAUC, 6576

\reference{} Clarke, T. E., Blaes, O., \& Tremaine, S. 1994, \aj, 107, 1873

\reference{} Dasgupta, A. K. 1979, \apss, 63, 517

\reference{} Dermer, C. D. 1996, in AIP Conf. Proc. 384, Gamma-Ray
   Bursts, ed. C. Kouveliotou, M. S. Briggs, \& G. J.  Fishman (New
   York: AIP), 744

\reference{} Fenimore, E. E., et al.\ 1993, \nat, 366, 40

\reference{} Fishman, G. J., et al.\ 1989, in Proc. of the Gamma Ray
   Observatory Science Workshop, ed. W. N. Johnson (Greenbelt:
   NASA/GSFC), 39

\reference{} Graziani, C., \& Lamb, D. Q. 1996, in AIP Conf. Proc. 384,
   Gamma-Ray Bursts, ed. C. Kouveliotou, M. S. Briggs, \& G. J.
   Fishman (New York: AIP), 382

\reference{} Greiner, J., et al.\ 1995, \aap, 302, 121

\reference{} Grindlay, J. E., \& Fazio, G. G. 1974, \apj, 187, L93

\reference{} Hanlon, L. O., et al.\ 1994, \aap, 285, 161

\reference{} Hanlon, L. O., et al.\ 1995, \aap, 296, L41

\reference{} Harrison, T. E., Webber, W. R., \& McNamara, B. J. 1995,
   \aj, 110, 2216

\reference{} Harrison, T. E., et al.\ 1995, \aap, 297, 465

\reference{} Hartmann, D. H., \& Blumenthal, G. R.\ 1989, \apj, 342, 521

\reference{} Hartmann, D. H., \&  Epstein, R. I.\ 1989, \apj, 346, 960

\reference{} Hartmann, D. H., Linder, E. V., \&  Blumenthal, G. R. 1991, 
   \apj, 367, 186

\reference{} Hartmann, D. H., et al.\ 1994, in AIP Conf. Proc. 307,
   Gamma-Ray Bursts, ed. G. J. Fishman, J. J. Brainerd, \& K. Hurley
   (New York: AIP), 127

\reference{} Higdon, J. C., \& Lingenfelter, R. E.\ 1990, ARA\&A, 28,
   401

\reference{} Horack, J. M., et al.\ 1994, \apj, 429, 319

\reference{} Howard, S., et al.\ 1993, in AIP Conf. Proc. 280, Compton
   Gamma Ray Observatory, ed. M. Friedlander, N. Gehrels, \&
   D. J. Macomb (New York: AIP), 793

\reference{} Hurley, K.\ 1982, in AIP Conf. Proc. 77, Gamma-Ray
   Transients and Related Astrophysical Phenomena, ed. R. E.
   Lingenfelter, H. S. Hudson, \& D. M. Worrall (New York: AIP), 85

\reference{} Hurley, K.\ 1986, in AIP Conf. Proc. 141, Gamma-Ray
   Bursts, ed. E. P. Liang, \& V. Petrosian (New York: AIP), 1

\reference{} Hurley, K., et al.\ 1992, \aaps, 92, 401

\reference{} Hurley, K., et al.\ 1994, in AIP Conf. Proc. 307,
   Gamma-Ray Bursts, ed. G. J. Fishman, J. J. Brainerd, \& K. Hurley
   (New York: AIP), 27

\reference{} Hurley, K., et al.\ 1996, in AIP Conf. Proc. 384,
   Gamma-Ray Bursts, ed. C. Kouveliotou, M. S. Briggs, \& G. J.
   Fishman (New York: AIP), 422

\reference{} Hurley, K., et al.\ 1997, \apj, 479, L113

\reference{} Katz, J. I. 1993, in AIP Conf. Proc. 280, Compton
   Gamma Ray Observatory, ed. M. Friedlander, N. Gehrels, \&
   D. J. Macomb (New York: AIP), 1090

\reference{} Kippen, R. M., et al.\ 1993a, in Proc. 23rd Intl. Cos.
   Ray Conf. (Calgary, Canada), 1, 85

\reference{} Kippen, R. M., et al.\ 1993b, in AIP Conf. Proc. 280, Compton
   Gamma Ray Observatory, ed. M. Friedlander, N. Gehrels, \&
   D. J. Macomb (New York: AIP), 823

\reference{} Kippen, R. M., et al.\ 1994a, IAUC, 5937

\reference{} Kippen, R. M., et al.\ 1994b, IAUC, 5943

\reference{} Kippen, R. M., et al.\ 1994c, in AIP Conf. Proc. 307,
   Gamma-Ray Bursts, ed. G. J. Fishman, J. J. Brainerd, \& K. Hurley
   (New York: AIP), 418

\reference{} Kippen, R. M., et al.\ 1995a, Adv. Space Res., 15, No. 5, 139

\reference{} Kippen, R. M., et al.\ 1995b, \aap, 293, L5

\reference{} Kippen, R. M., et al.\ 1995c, \apss, 231, 231

\reference{} Kippen, R. M., et al.\ 1995d, Ann. N.Y. Acad. Sci., 759, 425

\reference{} Kippen, R. M., et al.\ 1995e, in Proc. 24th Intl. Cos.
   Ray Conf. (Rome, Italy), 2, 61

\reference{} Kippen, R. M., et al.\ 1996a, in AIP Conf. Proc. 384,
   Gamma-Ray Bursts, ed. C. Kouveliotou, M. S. Briggs, \& G. J.
   Fishman (New York: AIP), 197

\reference{} Kippen, R. M., et al.\ 1996b, in AIP Conf. Proc. 384,
   Gamma-Ray Bursts, ed. C. Kouveliotou, M. S. Briggs, \& G. J.
   Fishman (New York: AIP), 436

\reference{} Kippen, R. M., et al.\ 1997, Adv. Space Res., in press

\reference{} Kolatt, T., \& Piran, T. 1996, \apj, 467, L41

\reference{} Kouveliotou, C., et al.\ 1996, in AIP Conf. Proc. 384,
   Gamma-Ray Bursts, ed. C. Kouveliotou, M. S. Briggs, \& G. J.
   Fishman (New York: AIP), 42

\reference{} Mallozzi, R. S., et al.\ 1993, in AIP Conf. Proc. 280,
   Compton Gamma Ray Observatory, ed. M. Friedlander, N. Gehrels, \&
   D. J. Macomb (New York: AIP), 1122

\reference{} Marani, G. F., Nemiroff, R. J., Norris, J. P., \&
   Bonnell, J. T. 1997, \apj, 474, 576

\reference{} McNamara, B. J., Harrison, T. E., \& Williams, C. 1995, \apj, 
   452, L25

\reference{} McNamara, B. J., et al.\ 1996, \apjs, 103, 173

\reference{} Meegan, C. A., et al.\ 1992, \nat, 355, 143

\reference{} Meegan, C. A., et al.\ 1995, \apj, 446, L15

\reference{} Meegan, C. A., et al.\ 1996, \apjs, 106, 65

\reference{} M\'{e}sz\'{a}ros, P., \& Rees, M. J. 1992, \apj, 397, 570

\reference{} Maoz, E. 1993, \apj, 414, 877

\reference{} Narayan, R., Paczy\'{n}ski, B., \& Piran, T. 1992, \apj,
   395, L83

\reference{} Nemiroff, R. J., Marani, G. F., \& Cebral, J. R. 1994,in
   AIP Conf. Proc. 307, Gamma-Ray Bursts, ed. G. J. Fishman,
   J. J. Brainerd, \& K. Hurley (New York: AIP), 137

\reference{} Paczy\'{n}ski, B. 1990, \apj, 348, 485

\reference{} van~Paradijs, J., et al.\ 1997, \nat, 386, 686

\reference{} Pendleton, G. N., et al.\ 1996, \apj, 464, 606

\reference{} Quashnock, J., \& Lamb, D. Q. 1993, \mnras, 265, 59p

\reference{} Ryan, J., Kippen, M., \& Varendorff, M. 1993, IAUC, 5702

\reference{} Ryan, J. M., et al.\ 1994a, IAUC, 5950

\reference{} Ryan, J. M., et al.\ 1994b, \apj, 422, L67

\reference{} Schaefer, B. E., \& Cline, T. L. 1985, \apj, 289, 490

\reference{} Schartel, N., Andernach, H., \& Greiner, J. 1997, \aap,
   in press

\reference{} Sch\"onfelder, V., et al.\  1991, IAUC, 5369

\reference{} Sch\"onfelder, V., et al.\  1993, \apjs, 86, 629

\reference{} Stacy, G. J., et al.\  1996, \aaps, 120, 691

\reference{} Tegmark, M., et al.\ 1996, \apj, 466, 757

\reference{} Varendorff, M. G., et al.\ 1992, in AIP Conf. Proc. 265,
   Gamma-Ray Bursts, ed. W. S. Paciesas, \& G. J. Fishman (New York: AIP), 77

\reference{} Varendorff, M. G., et al.\ 1993, in Proc. 23rd Intl. Cos.
   Ray Conf. (Calgary, Canada), 1, 81

\reference{} Vrba, F. J. 1996, in AIP Conf. Proc. 384, Gamma-Ray
   Bursts, ed. C. Kouveliotou, M. S. Briggs, \& G. J.  Fishman (New
   York: AIP), 565

\reference{} Wang, V. C., \& Lingenfelter, R. E. 1993, \apj, 416, L13

\reference{} Webber, W. R., et al.\ 1995, \aj, 110, 733

\reference{} White, R. S. 1993, \apss, 208, 301

\reference{} White, R. S. 1994, in AIP Conf. Proc. 307, Gamma-Ray
   Bursts, ed. G. J. Fishman, J. J. Brainerd, \& K. Hurley (New York:
   AIP), 620

\reference{} Winkler, C., et al.\ 1986, Adv. Space Res., 6, No. 4, 113

\reference{} Winkler, C., et al.\ 1992, \aap, 255, L9

\reference{} Winkler, C., et al.\ 1992, in AIP Conf. Proc. 265,
   Gamma-Ray Bursts, ed. W. S. Paciesas, \& G. J. Fishman (New York: AIP), 22

\reference{} Winkler, C., et al.\ 1993, in AIP Conf. Proc. 280,
   Compton Gamma Ray Observatory, ed. M. Friedlander, N. Gehrels, \&
   D. J. Macomb (New York: AIP), 845

\reference{} Winkler, C., et al.\ 1995, \aap, 302, 765

\end{references}
\end{document}